\documentclass{article}
\usepackage[utf8]{inputenc}
\usepackage{amsmath}
\usepackage{amssymb}
\usepackage{amsthm}
\usepackage{mathtools}
\usepackage{makecell}
\usepackage{subcaption}
\usepackage{xcolor}
\usepackage{braket}
\usepackage{hyperref}
\usepackage{graphicx,caption}
\usepackage[backend=biber,style=alphabetic,sorting=ynt]{biblatex}
\usepackage{authblk}

\usepackage{lineno}

\title{Efficient and fail-safe quantum algorithm\\ for the transport equation}
\author{Merel A. Schalkers\footnote{Corresponding author: \url{m.a.schalkers@tudelft.nl}} \and
        Matthias Möller}
\affil{Department of Applied Mathematics,\\ Delft University of Technology, The Netherlands}
\date{November 2022}

\begin{document}

\maketitle

\begin{abstract}
In this paper we present a scalable algorithm for fault-tolerant quantum computers for solving the transport equation in two and three spatial dimensions for variable grid sizes and discrete velocities, where the object walls are aligned with the Cartesian grid, the relative difference of velocities in each dimension is bounded by 1 and the total simulated time is dependent on the discrete velocities chosen. 
We provide detailed descriptions and complexity analyses of all steps of our quantum transport method (QTM) and present numerical results for 2D flows generated in Qiskit as a proof of concept.

Our QTM is based on a novel streaming approach which leads to a reduction in the amount of CNOT gates required in comparison to state-of-the-art quantum streaming methods.

As a second highlight of this paper we present a novel object encoding method, that reduces the complexity of the amount of CNOT gates required to encode walls, which now becomes independent of the size of the wall. Finally we present a novel quantum encoding of the particles' discrete velocities that enables a linear speed-up in the costs of reflecting the velocity of a particle, which now becomes independent of the amount of velocities encoded.

Our main contribution consists of a detailed description of a fail-safe implementation of a quantum algorithm for the reflection step of the transport equation that can be readily implemented on a physical quantum computer. This fail-safe implementation allows for a variety of initial conditions and particle velocities and leads to physically correct particle flow behavior around the walls, edges and corners of obstacles.

Combining these results we present a novel and fail-safe quantum algorithm for the transport equation that can be used for a multitude of flow configurations and leads to physically correct behavior. 

We finally show that our approach only requires $\mathcal{O}\left (  n_w n_g^2 + d n^v_t n_{v_\text{max}}^2 \right)$ CNOT gates, which is quadratic in the amount of qubits necessary to encode the grid and the amount of qubits necessary to encode the discrete velocities in a single spatial dimension. This complexity result makes our approach superior to state-of-the-art approaches known in the literature.

\end{abstract}

\section{Introduction}

Computational Fluid Dynamics (CFD) has become an indispensable third pillar in modern engineering sciences complementing theoretical and experimental analysis. Its broad applicability has led practitioners to constantly pushing the limits of numerical simulations for at least four decades. However, stagnation of Moore's law requires a radical rethinking of the way CFD codes and their underlying mathematical algorithms need to be designed in the future.

An emerging compute technology that has the potential to become a game changer for future CFD applications is quantum computing as it offers breakthrough solutions for the two major challenges of CFD today: large memory consumption and excessive need for computing power. The reader who is not familiar with the fundamental concepts of quantum computing is referred to Appendix~\ref{app:quantum computing} for a brief introduction into quantum bits and gates, entanglement of quantum states, the superposition principle, quantum parallelism and the way of extracting information from a quantum computer by means of measurement. A more extensive description of these concepts can be found in \cite{mikeandike,DeWolf}. 

In a nutshell, the advantage of quantum computers comes from their capability of encoding an exponentially large amount of data in linearly many quantum bits (qubits) and performing operations on all data simultaneously. This type of quantum parallelism is impossible with classical computers that either need to process data one by one, which results in an exponential growth in sequential run time, or duplicate the hardware resources and process multiple data in parallel, leading to an up to exponential growth in hardware resources. However, the advantage of quantum computing does not come for free. Extracting \emph{all} data from the quantum register requires an exponential amount of runs and measurements, thereby foiling any potential quantum advantage. The art of designing quantum algorithms with practical advantage consists of reducing the amount of necessary read-outs, e.g., by performing some post-processing of field data into scalar quantities of interest on the quantum computer itself.

CFD falls into the category of applications that have a good match with the capabilities and limitations of quantum computers, i.e., large amount of data to be worked with, high computational intensity, and, at the same time, users' primary interest in scalar-valued quantities of interest such as drag and lift coefficients, rather than the visual inspection of entire flow fields.

This has led early adopters of this emerging compute technology to explore the possibilities of speeding up CFD applications through the use of quantum computers. In what follows, we give a brief overview over the state-of-the-art in quantum CFD literature. 

In 2020 Gaitan published a quantum algorithm that can be used to solve the Navier-Stokes equations \cite{Gaitan2020}. In this article the author shows that there is a quantum speed-up for non-smooth flows and identifies a regime where the speed-up is quadratic over classical random algorithms.
Later, in 2021, Oz et al. adopted the algorithm presented in \cite{Gaitan2020} for solving partial differential equations to Burgers' equation \cite{Oz2021}. In this work the authors present solutions for flow problems with and without shock waves thereby achieving similar speed-up as for the Navier-Stokes equations. 
In a recent paper from 2022, Budinski suggests an alternative approach for solving the Navier-Stokes equations on a quantum computer by resorting to the stream-function vorticity formulation and adopting the lattice Boltzmann method \cite{Budinski2021}, thereby showing a proof of concept for yet another quantum method for solving the Navier-Stokes equations. 

A hybrid quantum classical reservoir computing model was suggested by Pfeffer et al. in 2022 \cite{Pfeffer2022}. With their method the authors are able to simulate nonlinear chaotic dynamics of Lorenz type models. They show that by using just a few highly entangled qubits they can achieve similar prediction and reconstruction capabilities as classical reservoirs using thousands of perceptrons. 

Another interesting approach that targets noisy intermediate quantum computers was proposed by Kyriienko et al. in \cite{Kyriienko2021}. In essence, a quantum neural network is trained to learn the solution values of a partial differential equation complemented by boundary and possibly initial values. The classical counterpart of this concept has become popular in recent years under the name physics-informed machine learning.

In 2021 Liu et al. presented a quantum algorithm for solving non-linear differential equations \cite{Liu2021}. The authors suggest to use Carleman linearization and subsequently perform time integration by the forward Euler method in combination with the quantum linear system algorithm by Harrow et al. \cite{Harrow2009} to find a solution. 

Very recently a paper was published that uses the Liouville equation to formulate a probability distribution of a flow field \cite{Succi2023} and proposes solving fluid flow problems on a quantum computer in that manner. 

Another approach that builds on the idea of distributions was proposed by Todorova and Steijl \cite{Todorova2020}, they resort to the transport equation and mimic the evolution of the particle distribution of highly rarified gases over time with a quantum circuit.

In the same year, Budinski suggested a quantum lattice Boltzmann method for the D1Q2 and D2Q5 lattice structure that includes a simplified collision term, but cannot model the specular reflection step \cite{Budinski2020}. The collision term is realized using the linear approximation of unitary approach \cite{Low2019}, which requires a measurement at the end of each time step to check whether the previous computations have been meaningful or ended up in a so-called `orthogonal state of no interest'. In the former case, the algorithm can proceed to the next time step. In the latter case, the entire computation must be quashed and the simulation must be started from scratch. Post-selection of valid results after (partial) measurement as used in the approximation of unitary approach is a common practice in quantum algorithms but it becomes problematic if used repeatedly within a time-stepping loop. It is obvious that the probability of failure increases in a multiplicative manner with the number of timesteps. Therefore the amount of timesteps that can be feasibly modeled using this approach highly depends on the amplitudes of the `orthogonal states of no interest' and is not realistically implementable for an interesting amount of timesteps, unless the probability of failure can be shown to be very small. Unfortunately, the paper \cite{Budinski2020} does not provide any bound on this probability. 

In the subsequent paper \cite{Budinski2021}, Budinski proposed an implementation of the specular reflection step using the quantum linear approximation of unitary approach, which brings the same problems as before. Furthermore, the paper does not provide a clear procedure for decomposing the required unitary into known quantum gates which hinders its straightforward implementation. 

In what follows, we propose a novel quantum algorithm for the transport equation that surpasses the aforementioned approach by Todorova and Steijl \cite{Todorova2020} in several key manners. First of all our quantum specular reflection step, even though seemingly more complicated, performs the correct reflection behavior in all corner cases. This is not the case for the original \textcolor{black}{transport equation method}, which shows incorrect reflection behavior for particles hitting the corners of an obstacle on a non-axis-aligned trajectory. On top of that, our specular reflection approach also ensures that the particles are set back into the flow domain before the start of a new timestep if they virtually traveled into an object. This also avoids incorrect behavior present in the original \textcolor{black}{transport equation method}.
Furthermore our approach for identifying the internal grid points of obstacles at which particles need to be reflected is more efficient than the method used by Todorova and Steijl. This leads to an additional polynomial speed-up.

We finally show that our quantum implementation of the streaming step outperforms the method of \cite{Todorova2020} in terms of the the amount of CNOT gates required to implement them. We intentionally choose for the latter as performance measure as they are either available as native gates or can be efficiently emulated. Hence, complexity estimates in terms of CNOT gates give a more realistic picture of the costs on a real quantum computer, whereas complexity estimates in terms of multi-controlled gates conceal the costs for decomposing multi-controlled gates into native ones.

Last but not least our method outperforms the current state-of-the-art approach in the encoding of the particles' velocity. We propose a novel encoding of the velocity vector, due to which velocities can be flipped with a single NOT operation performed on a single qubit, whereas before $n$ NOT operations were required to flip the direction of the velocity encoded using $n$ qubits \cite{Todorova2020}.

The rest of the paper is structured as follows. In Section \ref{sec:collisionless boltzmann equation} we give a brief introduction to the \textcolor{black}{transport} equation.
Section \ref{sec:quantum register set-up} shows how the data required for our \textcolor{black}{QTM} can be encoded efficiently in a quantum register. Here, we define the grid-set up, use of ancillae and the novel velocity-vector encoding that enables the flipping of the velocities being possible in constant time.
Subsequently, Section \ref{sec:efficient quantum convection operation} provides an efficient quantum incrementation (decrementation) method, which provides a significant speed-up when decomposed into native gates over the incrementation (decrementation) method used in other quantum implementations. We then show how this novel quantum incrementation step can be used to implement a quantum streaming operation. 
The fail-safe quantum specular reflection step is presented in Section \ref{sec:complete quantum specular reflection step}. Here, we provide an extensive and implementable method to avoid deviant behavior around the corners of objects while ensuring unitarity. We furthermore propose a new approach to ensure that the particles will be repositioned into the flow domain before the start of the next timestep. On top of that we design an efficient method to identify whether or not a grid point is located inside an internal obstacle.  
To wrap everything up and show the functionability of our approach Section \ref{sec:results} gives the result of preliminary simulation runs. 
Finally, Section \ref{sec:comparison to state of the art} compares the complexity of our approach to that of other known methods.
Appendix \ref{app:quantum computing} provides an introduction to the elements of quantum computing necessary to understand the presented \textcolor{black}{QTM} for the non-quantum expert. Appendix \ref{app:glossary} gives an overview of the parameters and terms used in this paper.

\section{The transport equation}\label{sec:collisionless boltzmann equation}
In statistical physics it is customary to describe the dynamics of gases by tracking the position $\mathbf{x}_a=(x^{1}_a,x^{2}_a,x^{3}_a)$ and momentum $\mathbf{p}_a=(p^{x}_a,p^{y}_a, p^{z}_a)$ of each single molecule $a=1,\dots,N$ in the $6N$-dimensional phase space over time. As it is impractical to work in phase space for large $N$ due to its enormous dimensionality, it is common practice to change to the six-dimensional $\mu$-space whose coordinates are $(x^{(1)},x^{(2)},x^{(3)},p^{(1)},p^{(2)},p^{(3)})$. Instead of tracking the evolution of individual molecules, in the $\mu$-space we resort to the distribution function $f(\mathbf{x},\mathbf{u},t)$ to describe the density of molecules at a position $\mathbf{x}=(x^{(1)},x^{(2)},x^{(3)})$ with velocity $\mathbf{u}=(u^{(1)},u^{(2)},u^{(3)})$ at time $t$. 

In this paper we use the transport equation to describe the evolution of the density of molecules through time. The transport equation is given by
\begin{equation}
  \frac{\partial f}{\partial t} + \mathbf{u}\cdot \nabla f = 0.
  \label{eq:collisionless_boltzmann}
\end{equation}
This equation describes the transport of particles with a given velocity field but neglects the interaction of particles which is much more difficult to realise with a quantum algorithm. 

We assume that the speeds are discrete, i.e., $u^{(i)}\in\mathcal{U}^{(i)}=\{u_1,\dots ,u_{N_{v_i}}\}$. For the scope of this paper we only consider velocities $\mathbf{u}=(u^{(1)},u^{(2)},u^{(3)})$, for which the relative difference in speed in the different dimensions is bounded by
\begin{equation}
    c_\text{rel} = \max \frac{|u^{(i)}|}{|u^{(j)}|}\leq 1.
\end{equation}
As detailed in Section \ref{ssec:class_reflection} this is not a conceptual limitation of our approach, but helps us to keep the number of corner cases to be considered separately small, which also improves the readability of the paper. 

\subsection{Grid definition and obstacle placement}\label{ssec:grid}
\begin{figure}[h]
\centering
\includegraphics[]{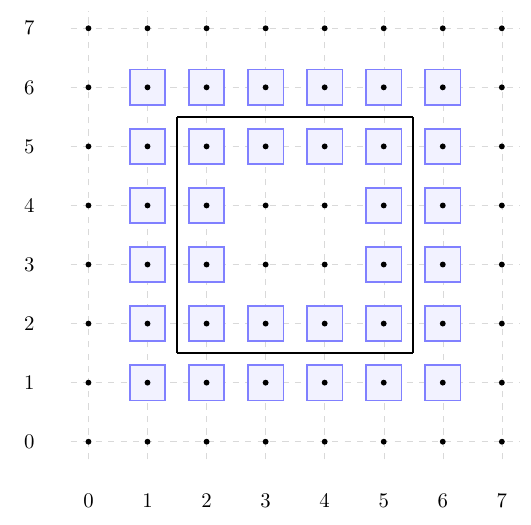}
    \caption{Illustration of an obstacle (black box) placed in a computational grid in two dimensions. The grid points (`$\cdot$' surrounded by blue boxes) surrounding the obstacle placement require special attention in our implementation. }
    \label{fig:grid}
\end{figure}

In order to keep track of the temporal evolution of the distribution function $f(\mathbf{x},\mathbf{u},t)$, we first define a computational grid with equidistant grid spacing in all two or three spatial dimensions and assign densities to the volumes centered around the grid points. The grid is set up in a straightforward fashion, and grid points can be identified using the Cartesian coordinates system. Obstacles are subsequently placed in between grid points as depicted in Figure~\ref{fig:grid}.

\subsection{Streaming}
In what follows we describe how to move the particles through space.
In the \textcolor{black}{transport} method not all particles move in each time step. 
Furthermore, when a particle moves it does not necessarily move in all dimensions in the same time step. The velocity of each particle consists of a speed in each spatial dimension. Whether or not a particle will move in a given dimension depends on the speed in that dimension and whether or not a particle traveling at that speed will reach the next grid point in the current time step. 
This is due to the fact that we choose the size $\Delta t_m$ of time step $m$ to be such that some particles make it to the next grid point, but none overshoot. 
To make this idea more precise we define $u_\text{max} = \max_k |u_k|$ and $u_\text{min} = \min_k |u_k|$, with $u_k \in \mathcal{U} \coloneqq \bigcup_{i=1}^d \mathcal{U}^{(i)}$ the set of possible speeds a particle can travel. Then, we can normalize the distance between the grid points $\Delta x$ to enforce it will always be equal to $1$. Subsequently, the first time step $\Delta t_0$ will have size $\frac{1}{u_\text{max}}$. All particles that travel with speed $\pm u_{\text{max}}$ in a dimension will move one grid point in the first time step in that dimension. To determine the size $\Delta t_m$ of any subsequent time step $m$, we keep track of the distances a particle traveling at any of the possible speeds $u_k\in\mathcal{U}$ would be removed from reaching the next grid point and use this to find the smallest time step necessary for any particle to reach the next grid point in any dimension.

To ensure that particles can travel no further than the neighboring grid points, and thereby adhere to the CFL (Courant-Friedrichs-Lewy) condition, we use a so called CFL counter. This counter keeps track of the distances the particles are located from the next grid point based on their speed in each dimension and thereby determines the size of the next time step. 

\subsubsection{CFL counter}\label{ssec:cfl_counter}
In what follows, let $c^m_k$ represent how large of a fraction from the current grid point to the next a particle with speed $u_k$ has to travel.

We use the following expression
\begin{equation}
    c^{m+1}_{k} = c^m_{k} + |u_k| \frac{\Delta t_m}{\Delta x},
\end{equation}
where $u_k \in \mathcal{U}$ and $m$ represents the number of time steps that have been taken so far. Furthermore, $\Delta x$ represents the equidistant spacing between the grid points as before.  
It then follows that we can express the size of the time step taken in each iteration as
\begin{equation}\label{eq:delta_t}
    \Delta t_m = \min_{k} \left ( \left [1-c^m_{k} \right ] \frac{\Delta x}{|u_k|}  \right ).
\end{equation}
In the above equation we minimize over $k$ to ensure that at least one speed reaches the next grid point, and none overshoots. 

In a classical \textcolor{black}{transport} method $c_k^{m}$ can be used to interpolate the results when the simulation terminates in a time step that is not a complete cycle. This is not possible in the quantum counterpart and thus if we terminate the simulation in a time step that is not a complete cycle, some small oscillations will occur unless a post-smoothing method is applied. We will address this issue in a forthcoming publication. In this paper we restrict ourselves to running the \textcolor{black}{QTM} algorithm for a total time $T = \sum_{m=0}^{N_t -1} \Delta t_m$ such that $\frac{T}{u_m} \in \mathbb{Z}$ $\forall m$. 

To complete the quantum \textcolor{black}{transport} method we need to describe the behavior when a particle impinges on an obstacle. 

\subsection{Reflection by an obstacle}\label{ssec:class_reflection}
When a particle impinges on an object its velocity gets reflected along the axis normal to that object. 
Notice that due to this method we are restricted to modeling objects whose walls are either parallel or perpendicular to each dimension.
In order to facilitate the correct reflections we need to keep track of when a particle has come into contact with a wall, this is done differently for different values of $c_\text{rel}$.

\begin{figure}
\centering
\includegraphics[]{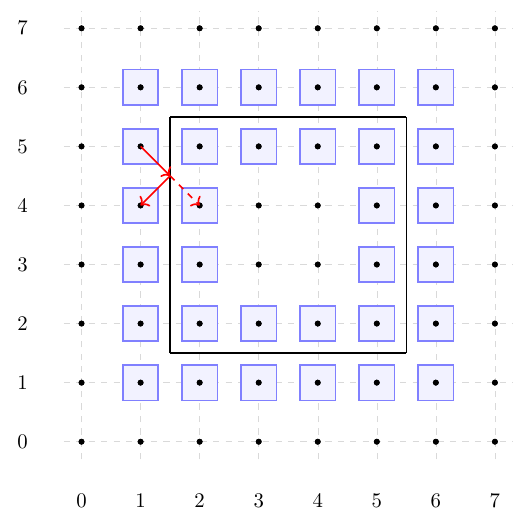}

    \caption{Illustration of an obstacle (black box) in the grid with the red arrow representing one possible specular reflection operation. The dashed arrow represents the trajectory had no reflection taken place. }
    \label{fig:specular reflection}
\end{figure}

We first consider the case $c_\text{rel} \leq 1$. In this case we know a particle has come into contact with an object if and only if it hits a grid point located in the first layer of the obstacle. We will refer to these grid points as the wall of the object. When a particle reaches a grid point in the wall of an object, its velocity normal to the wall gets reversed and the particle is moved one grid point outside of the object again in the direction normal to the wall(s) that just reflected it. Figure \ref{fig:specular reflection} gives an example of what such a reflection might look like.

\begin{figure}
\centering
\includegraphics[]{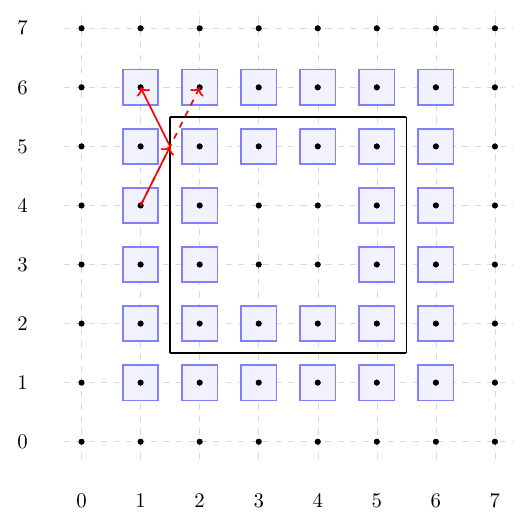}

    \caption{Illustration of an obstacle (black box) in the grid with the red arrow representing one possible specular reflection operation when $c_\text{rel} > 1$ holds. The dashed arrow represents the trajectory had no reflection taken place. }
    \label{fig:specular reflection c3}
\end{figure}

For $c_\text{rel} > 1$ the specular reflection step becomes more complicated, as grid points located outside the object can be reached by a particle indicating that the particle has come into contact with an object; See Figure \ref{fig:specular reflection c3}. Similarly multiple cases need to be taken into account to determine whether a particle hits a corner point of an object or the wall at a non-corner point. For simplicity we will restrict ourselves to the case $c_\text{rel} \leq 1$ for the scope of this paper, but we would like to remark that our method can be generalized to any value of $c_\text{rel}$ at the cost of taking into account more corner cases as will be detailed in a future publication. 

\begin{figure}
\centering
\includegraphics[scale=.5]{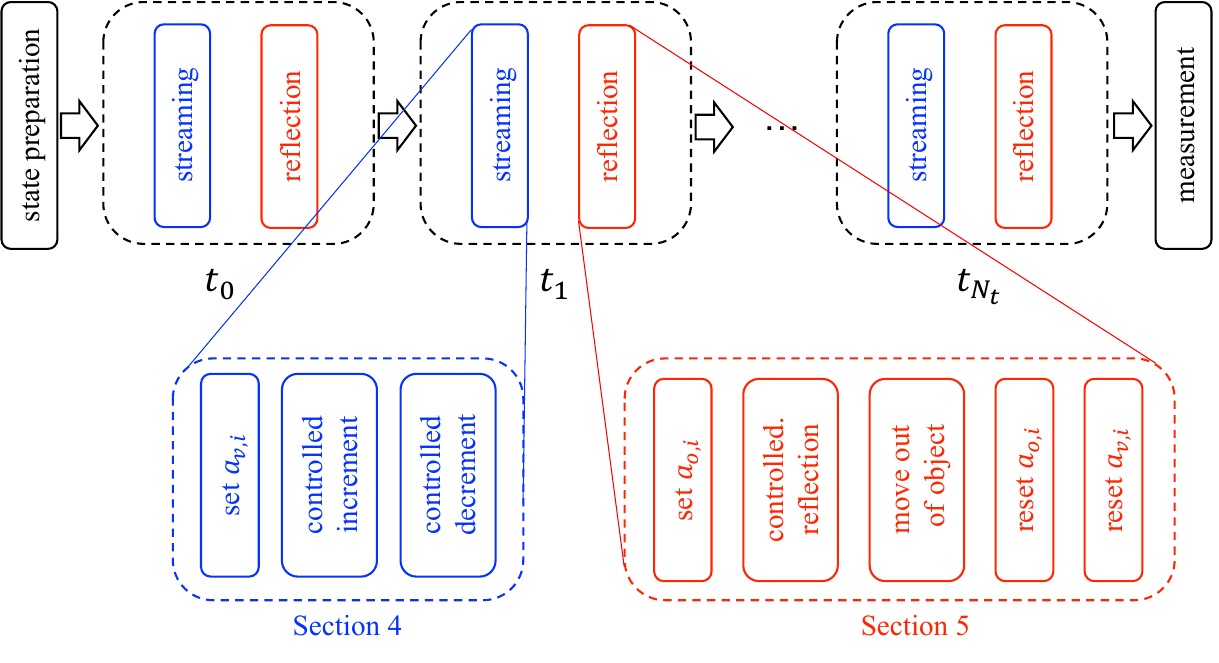}

    \caption{Schematic overview of the proposed \textcolor{black}{QTM}. }
    \label{fig:schematic overview}
\end{figure}

A schematic overview of the proposed \textcolor{black}{QTM} is given in Figure \ref{fig:schematic overview}.

\section{Quantum register set-up}\label{sec:quantum register set-up}
In order to implement our \textcolor{black}{QTM} approach we first need to define an encoding to give a physically interpretable meaning to quantum basis states. 
In our implementation we define a mapping, where each parameter is assigned to its own set of qubits, which combined form the total qubit register. 
We use the following encoding
\begin{equation}
    \ket{\underbrace{a_{n_a}\dots a_{1}}_{\mathrm{ancillae}} \overbrace{ g_{n_g} \dots g_{1} }^{\mathrm{position}} \underbrace{ v_{n_v} \dots v_1}_{\mathrm{velocity}}}.
\end{equation}

Here, the qubits $a_{n_a}, \dots, a_1$ form the ancillae, with $n_a = 4d-2$, where $d$ is the number of spatial dimensions we are modeling. The $g_{n_g} \dots g_1$ qubits form the positional qubits, with $n_g = \sum_{i=1}^{d}n_{g_i}$, where $n_{g_i}$ is the number of qubits required to number all grid points in the $i$-th spatial dimension. Finally, the qubits $v_{n_v} \dots v_1$ encode the velocity vector, with $n_v = \sum_{i=1}^{d}n_{v_i} $ where $n_{v_i}$ is the number of qubits required number the velocities of the $i$-th dimension. In summary, the total number of qubits required to realize our quantum \textcolor{black}{transport} method is
\begin{equation}
    a_n + n_g + n_v = 4d-2 + \sum_{i=1}^d (n_{g_i}+n_{v_i}),
\end{equation}
which means that a moderate number of a few dozens to hundred fault-tolerant qubits suffice to enable the solution of two and three-dimensional problems.

The details of the mapping of the ancillae, positional and velocity vectors will be explained more in depth in the following sections.

\subsection{Efficient mapping of velocity vector}
One of the speed-ups our algorithm provides compared to the state-of-the-art is due to our mapping of the velocity vector. 
In each spatial dimension we are working with a predetermined amount of discrete velocities $N_{v_i}$. 
For simplicity we will at first only consider the velocity encoding in the one-dimensional case, and call the amount of discrete velocities $N_{v}$. Note that this approach trivially extends to the multi-dimensional case.
As $N_v$ is the number of discrete velocities required, we need $n_{v} = \lceil\log_2\left(N_v\right)\rceil$ qubits to encode all velocities. \\
For our \textcolor{black}{QTM} we restrict ourselves to the case that the set of speeds $\mathcal{U}$ consists of positive and negative values, and for each $u\in \mathcal{U}$ both $|u|\in \mathcal{U}$ and $-|u|\in \mathcal{U}$ will hold. We now define $\mathcal{U}_o = [-u_{\max}, \dots , u_{\max} ]$, the ordered list of speeds from most negative to most positive. 
Let $u_{\min} = \min_k \; |u_k| $ as before and let $\Delta u $ be the distance between the neighboring speeds in $\mathcal{U}_o$, for simplicity we will assume $\Delta u$ is constant between all indices so $\Delta u=\frac{2u_{\max}}{N_v}$. However, we are not restricted to this simplification.  \\

We propose the mapping
\begin{align}
\ket{v_{n_v} \dots v_1} = 
    \begin{bmatrix}
    -u_{\min} \phantom{\, -\Delta u}\\
    -u_{\min} -\Delta u\\
    \vdots \\
    -u_{\max} \phantom{\, -\Delta u}\\
    \phantom{-}u_{\min} \phantom{\, -\Delta u}\\
    \phantom{\,-}u_{\min} +\Delta u  \\
    \vdots \\
    \phantom{-}u_{\max} \phantom{\, +\Delta u}\\
    \end{bmatrix},\label{eq:our_v}
\end{align}
of the discrete velocities to the quantum state.
The motivation of this velocity mapping is that we can flip the direction of the velocity by flipping only one qubit. This can be seen by noting that the distance between the index of velocity $-|u_k|$ and $|u_k|$ in $\mathcal{U}_o$ is precisely $\frac{N_v}{2} = 2^{n_v-1}$, and so we can flip between these two velocities by flipping the most significant qubit which encodes the sign of the velocity vector.

This means that we can write the quantum register encoding the velocity in the one dimensional case as
\begin{equation}
   \ket{v_{\text{dir}} v_{n_v -1} \dots v_1}.
\end{equation}
In the above expression the first qubit encodes the direction of the velocity (positive or negative in the given dimension), and the next $n_v-1$ qubits encode the magnitude of the velocity. We define $\ket{00 \dots 0}$ to encode $-u_{\min}$, $\ket{00 \dots 1}$ to encode $-u_{\min} - \Delta u$ and $\ket{11 \dots 1}$ to represent $u_{\max}$ etc. It can easily be seen that this leads to a velocity encoding as given in Equation \eqref{eq:our_v}.

\paragraph{Example}
\emph{We give an example of our velocity encoding to show that flipping the direction can be established by flipping only the most significant qubit. Let us consider the 1-dimensional case and say we have $N_v = 8$. Now let $u_1 = u_{\min}$, $u_2 = u_{\min} + \Delta u$ and $u_3 = u_{\min} + 2\Delta u$ etc., then our velocity encoding gives $$\ket{v} = \begin{bmatrix}
-u_1 \\ -u_2\\ -u_3 \\ -u_4\\ \phantom{-}u_1 \\ \phantom{-}u_2\\ \phantom{-}u_3 \\ \phantom{-}u_4 
\end{bmatrix}.$$ Say our particle is traveling with velocity $u_2$, then $\ket{v} = \ket{101}$. Now flipping the most significant qubit gives $\ket{v} = \ket{001}$ which maps to $-u_2$ as required. }\\

Extending the above approach to the multidimensional case, the quantum register for the $d$-dimensional velocity becomes 
\begin{equation}
\ket{v_{\text{dir},d} \dots v_{\text{dir},1} v^d_{n_{v_d}} \dots v^d_{1} v^{d-1}_{n_{v_{d-1}}} \dots v^{d-1}_{1} \dots v^1_{n_{v_1}} \dots v^1_{1}} . 
\end{equation}

Former encodings of the velocity vector for quantum \textcolor{black}{transport} methods were set-up such that in order to flip the velocity of a particle, all qubits encoding the velocity needed to be flipped \cite{Todorova2020}. Our new encoding saves $\mathcal{O}\left ( n \right )$ qubit flips per reflection. Notice that since reflections are usually implemented in multi-control fashion, as described in Section \ref{sec:complete quantum specular reflection step}, this ends up in saving $\mathcal{O}\left (n \right )$ computationally costly operations per reflection. In Section \ref{sec:comparison to state of the art} we give an in depth complexity analysis and comparison of the different methods.

\subsection{Mapping of grid point locations onto qubit states}
The mapping of the grid point locations onto qubit states is rather straightforward. As before, let $\ket{g_{n_g} \dots g_1}$ represent the state of the qubits encoding the grid point location of the particles. Then, we can write the qubits encoding the location more detailed as
\begin{equation}
\ket{g^d_{n_{g_d}} \dots g^d_{1} g^{d-1}_{n_{g_{d-1}}} \dots g^{d-1}_{1} \dots  g^1_{n_{g_1}} \dots g^1_{1}},
\end{equation}
where $g^i_{n_{g_i}} \dots g^i_{1} $ encodes the $i$-th dimension of the location of grid points by representing the binary value of the location. The set-up of the grid is rather straightforward with different grid-points in each dimension and the particles being able to move one step forward and backward in each time step and in each dimension.

\subsection{Ancillae}
The ancillae are the last of the register expressed in Equation \eqref{eq:our_v} and are used within the computation only. 
Each dimension requires one ancilla to keep track of which velocities will be streamed and one ancilla to keep track of the resetting
during the reflection step, furthermore we require $2(d-1)$ ancillae to implement a quantum comparator method that will be used in the reflection step, leading to a total of $4d-2$ ancillae required throughout the computation.
The exact set-up and utilization of the ancillae will be described in Sections \ref{sec:efficient quantum convection operation} and \ref{sec:complete quantum specular reflection step}.

\section{Efficient quantum streaming operation}\label{sec:efficient quantum convection operation}
As described in Section \ref{sec:collisionless boltzmann equation} the main ingredients of the quantum \textcolor{black}{transport} method are the streaming and the reflection operations. In this section we will introduce a novel streaming operation based on the Quantum Draper Adder \cite{Draper1998}, which we specialize to an efficient quantum incrementation (decrementation) procedure that is cheaper than methods currently in use \cite{Douglas2009,Fillion-Gourdea2017, Fillion-Gourdea2018,Todorova2020}; see Section \ref{ssec:comp_soa_q}. \\

\subsection{Efficient quantum incrementation (decrementation)}\label{ssec:efficient quantum incr}
An increment (decrement) operation takes a quantum state $\ket{j}$ to the state $\ket{j+1}$ $\left (\ket{j-1} \right )$. This operation is cyclic, meaning that $\ket{2^n-1}$ ($\ket{0}$) gets incremented (decremented) to $\ket{0}$ $\left (\ket{2^n-1} \right )$.  
Let $U_\text{inc}$ ($U_\text{dec}$) express the unitary that increases (decreases) the $n$ qubit state $\ket{j}$, this gives
\begin{equation}
    U_\text{inc} \ket{j} = \ket{j+1},
\end{equation}
\begin{equation}
    U_\text{dec} \ket{j} = \ket{j-1}.
\end{equation}
This quantum primitive is used in many different quantum algorithm and fields such as Quantum Random Walks and Quantum Computational Fluid Dynamics, to name just a few. In the literature this primitive is typically implemented by cascading multi-controlled NOT operations \cite{Douglas2009,Fillion-Gourdea2017, Fillion-Gourdea2018,Todorova2020, Budinski2021}. While this implementation looks elegant on paper, decomposing multi-controlled NOT operations into gates that are native to quantum computers, i.e., single-controlled NOT gates, will lead to a significant increase of the circuit depth. 

In this paper we provide an alternative method leading to a quantum streaming operation which can be implemented more efficiently on real-world quantum computers. The method is inspired by the Quantum Draper Adder (QDA) \cite{Draper1998} and uses the same principle, but in contrast to the regular Drapper adder that computes $\ket{a + b}$, where $a$ and $b$ are natural numbers encoded in a quantum register, in our implementation the phase shift operations are not controlled as the addition by $b=1$ is always known beforehand. As we moreover want our operation to be cyclic, no qubit for holding a potential carry over value is required. We are, to the best of our knowledge, the first to use such a QDA inspired approach for the quantum incrementer (decrementer) and in Section \ref{ssec:complexity incrementation} we show that it leads to a significant reduction in the amount of CNOT gates required to run the algorithm. \\

Our method can be expressed by the circuit given in Figure \ref{fig:streaming_no_anc}. 
Let $\ket{j}$ be the basis state that we wish to increment (decrement). We claim that
\begin{equation} \label{eq:our_adder}
    \ket{j+1} = (QFT)^\dagger U_{P,+} (QFT) \ket{j},
\end{equation} and define 
\begin{equation} \label{eq:def our_adder}
    U_\text{inc}= (QFT)^\dagger U_{P,+} (QFT),
\end{equation} where QFT stands for Quantum Fourier Transform and $U_{P,+}$ will be defined below.
We will now show that \eqref{eq:our_adder} indeed holds. First we remark that
\begin{equation}
    QFT \ket{j} = \frac{1}{\sqrt{N}}\sum_{k=0}^{N-1} \omega_N^{kj}\ket{k} = \frac{1}{\sqrt{N}} \sum_{k=0}^{N-1} e^{\frac{2 i \pi k j}{N}}\ket{k},
\end{equation}
holds by definition.
Then the circuit $U_{P,+}$ consists of one phase shift gate\footnote{The single qubit phase shift gate can be expressed in matrix form as $P\left ( \theta \right ) = \begin{bmatrix}1 & 0 \\ 0 & e^{i\theta} \end{bmatrix}$.  } per qubit with an angle determined by the qubit number of the qubit to which it is applied. We apply the phase shift gate $P(\theta)$ to the $j$-th qubit\footnote{In this section, for simplicity, we index the qubits ranging from 0 to $n-1$.} with angle
\begin{equation}\label{eq:theta_i}
    \theta_j=\frac{\pi}{2^{n-1-j}} =\frac{\pi 2^{j+1}}{N}.
\end{equation}

So $P(\theta_j)$ gets applied to qubit $j$ and adds amplitude $e^{\frac{i2\pi 2^j}{N}}$ to each basis state for which qubit $j$ is in the state $1$. This means that if we apply the operation $P(\theta_j)$ to each qubit for the basis state $\ket{k}$ it gets multiplied by a factor $e^{\frac{2\pi i k}{N}}$. 

And so we get
\begin{equation}
    U_{P,+} (QFT) \ket{j} = \frac{1}{\sqrt{N}} \sum_{k=0}^{N-1} e^{\frac{2 i \pi k j}{N}}e^{\frac{2\pi i k}{N}} \ket{k} = \frac{1}{\sqrt{N}} \sum_{k=0}^{N-1} e^{\frac{2 i \pi k (j+1)}{N}}\ket{k} .
\end{equation}

It then directly follows that
\begin{equation}
    (QFT)^\dagger U_{P,+} (QFT) \ket{j} = QFT^\dagger \frac{1}{\sqrt{N}} \sum_{k=0}^{N-1} e^{\frac{2 i \pi k (j+1)}{N}}\ket{k}  = \ket{j+1}.
\end{equation}
In other words our proposed algorithm increases each basis state by $1$, whereby the periodic property of $e^{i\theta}$ ensures that the so-defined increase operation is cyclic. Naturally the decrease operation becomes
\begin{equation}
    \ket{j} =  \left ( (QFT)^\dagger U_{P,+} (QFT) \right )^\dagger \ket{j+1} = (QFT)^\dagger U_{P,+}^\dagger (QFT) \ket{j+1}.
\end{equation}
Since $U_{P,+}$ consists of a layer of phase shift gates taking the conjugate transpose is equal to shifting the angles to their negative counterparts. So for $U_{P,+}^\dagger$ we apply the phase shift gate $P(-\theta_j)$ to the $j$-th qubit with angle $-\theta_j$, where $\theta_j$ is as given in \eqref{eq:theta_i}.
For simplicity we define $U_{P,-} = U_{P,+}^\dagger$. It then follows that
\begin{equation}
    U_{\text{dec}} = (QFT)^\dagger U_{P,-} (QFT).
\end{equation}

\subsection{Streaming step}\label{ssec:convection step}
In the streaming step particles migrate from one discrete point in space to another. 
In each time step the particles that travel at a discrete speed such that they reach the next grid point in the current time step get incremented (or decremented) one position in space. The list of speeds for which particles traveling at that specific speed need to be incremented (decremented) at a particular time step is pregenerated by the CFL counter as described in Section \ref{ssec:cfl_counter}. Then, controlled on their speeds, the particles are incremented (decremented) one position at a time. This means that we implement the quantum incrementation (decrementation) method as described in the last section in a controlled fashion as will be detailed below.  \\

First of all we notice that we do not have to control the entire incrementation (decrementation) primitive from Section \ref{ssec:efficient quantum incr}. Since the incrementation (decrementation) step consists of the QFT followed by a layer of phase shift gates followed by QFT$^\dagger$, it suffices to only control the layer of phase shift gates on the speed of the particles, as the QFT and QFT$^\dagger$ operations naturally cancel out each other. Furthermore, since the incrementation and decrementation step are performed directly after each other on the same qubits, we do not have to perform a QFT$^\dagger$ operation after the phase shift gates of the incrementation step, as this would cancel out with the QFT operation at the start of the decrementation step.  \\

Second, we notice that if we make use of an ancilla we can perform the incrementation and decrementation operations for all speeds that take a step in the current time step using a single operation in a given dimension $i$. 

To this end, let $|u_k|$ be (one of) the absolute values of the speeds for which the streaming step is to be performed at a certain point in the algorithm. Then in each dimension $i$ we need to increment the particles traveling at speed $u^{(i)} = \pm |u_k|$. Because of the way the binary encodings of $u_k$ and $-u_k$ are related in each dimension, we first perform a controlled-NOT operation between the $n_{v_i}-1$ qubits determining the absolute value of the velocity and an ancilla qubit $a_{v,i}$. If there are multiple values $k$ such that particles traveling at speed $\pm u_k$ need to be incremented in the given time step, this process is applied to all such $k$. This means that we simply flip the value of the ancilla qubit $a_{v,i}$ to 1 for all the registers of the particles traveling at a speed such that they should reach the next grid point in dimension $i$ in the current time step. We then use the ancilla $a_{v,i}$ in combination with the directional velocity qubit $v_{\text{dir},i}$ to increase or decrease the position of the correct particles in space in each dimension $i$. 
In practice, the streaming step using the ancilla qubits $a_{v,i}$ should be implemented as it leads to a more efficient algorithm, due to the fact that we now perform the incrementation (decrementation) primitive for all the particles taking a step in the given time step using a single operation. 

\begin{figure}
    \centering
    \includegraphics{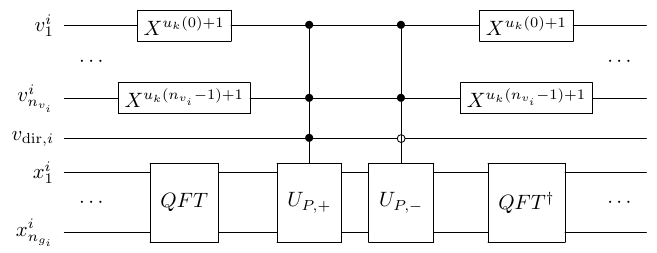}
    \caption{Circuit representation of the proposed streaming step in dimension $i$ for one speed $|u_k|$ without extra ancilla qubits.}
    \label{fig:streaming_no_anc}
\end{figure}

\begin{figure}
    \centering
    \includegraphics{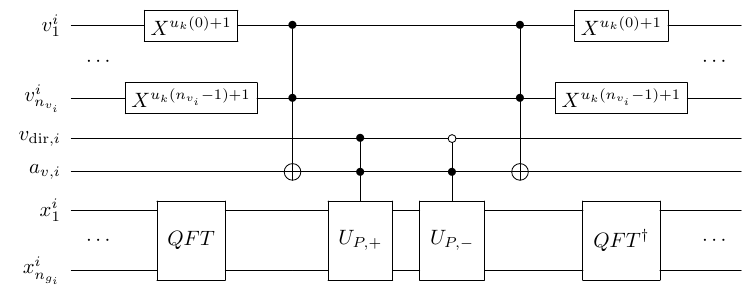}
    \caption{Circuit representation of the proposed streaming step in dimension $i$ for one speed $|u_k|$ with ancilla qubit. Note that here we reset the $a_{v,i}$ ancilla after the streaming step. In the proposed algorithm this ancilla will only be reset after the reflection operation has been performed. }
    \label{fig:streaming_anc}
\end{figure}

\begin{figure}
    \centering
    \includegraphics[scale = .75]{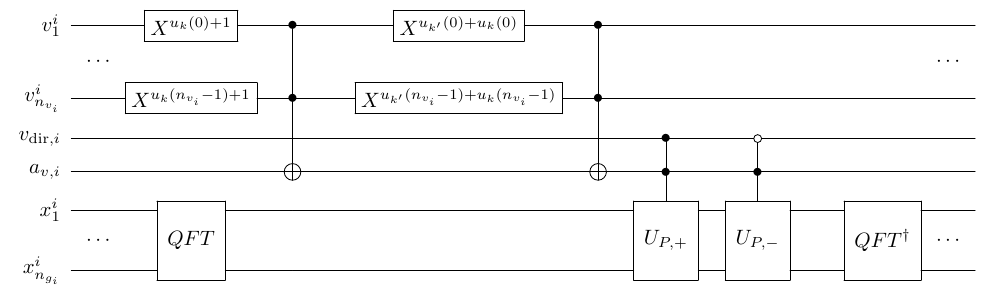}
    \caption{Circuit representation of the proposed streaming step in dimension $i$ for two speeds $|u_k|$ and $|u_{k^\prime}|$ with ancilla qubit. Note how the proposed ancilla qubit allows to perform the streaming $U_{P,+}$ for both speeds in one operation. Here we do not reset the ancilla qubit $a_{v,i}$ after the streaming step, as in the proposed algorithm this ancilla will only be reset after the reflection operation has been performed.}
    \label{fig:streaming_anc two}
\end{figure}

Let ${u_{k}(n)}$ represent the value of the $n$-th bit when representing the index of the speed $u_{k}$ in the velocity vector\footnote{For example, assume we want to control on the speed being equal to $u_2$, let $\ket{u_2} = \ket{v_\text{dir}v_m \dots v_0} = \ket{10010}$. Then ${u_{2}(0)} = 0$ and ${u_{2}(1)} = 1$ etc. }. Then, we perform the operation $X^{u_k(l-1)+1}$ on each qubit $v^i_l \in \{v^i_{1}, \dots, v^i_{n_{v_i}}\}$ followed by the multi-controlled NOT operation between the $v^i_{1} \dots v^i_{n_{v_i}}$ qubits and the $a_{v,i}$ qubit for all speeds $|u_k|$ that take a step in the given time step, before performing the controlled streaming operations. Notice that we need to make sure to apply the $X^{u_k(l-1)+1}$ operations again after performing the multi-controlled NOT, to reset the speeds to their original states before we apply the $X^{u_{k^\prime}(l-1)+1}$ associated with the next speed $u_{k^\prime}$ for which the particles need to be streamed in this timestep. Furthermore, notice that we can simply combine these subsequent operations and get $X^{u_k(l-1)+1+u_{k^\prime}(l-1)+1} = X^{u_k(l-1)+u_{k^\prime}(l-1)}$ to be applied before the multi-controlled NOT operation to set the ancilla $a_{v,i}$ for the speed $u_{k^\prime}$. Figure \ref{fig:streaming_no_anc} shows the circuit of the streaming step without the ancilla $a_{v,i}$. Figure \ref{fig:streaming_anc} shows what the circuit of the streaming step with the ancilla $a_{v,i}$ looks like when streaming the particles with speed $u_k$ in dimension $i$. Figure \ref{fig:streaming_anc two} shows how using the ancilla qubit $a_{v,i}$ enables streaming the particles with several (in this example two) speeds $u_k$ and $u_{k^\prime}$ in dimension $i$ using a single $U_{p,+}$ and $U_{p,-}$ operation. In Figure \ref{fig:streaming_anc two} we do not reset the ancilla qubit $a_{v,i}$ after performing the streaming step, as in practice these will only get reset after the specular reflection step of the algorithm has been performed. 

In the above we have presented the streaming algorithm for a single dimension $i$. When working with multiple dimensions we perform the exact same operations in each dimension. Note how the streaming of the particle in the different dimensions can be applied to the qubits simultaneously, as different qubits are involved in the streaming step for each dimension.

\section{Quantum specular reflection step}\label{sec:complete quantum specular reflection step}
After the completion of the streaming operation, particles that come into contact with an obstacle and have virtually moved into it have to change their travel path; See Section \ref{ssec:class_reflection}. 

Here, we propose a novel and fail-safe approach to the quantum reflection operation.  
First, the particles that virtually traveled into the obstacle have their velocity direction reversed in the direction normal to the wall encountered. 
Afterwards, these particles are placed back into the flow domain. As a result of both operations, the particle is located in the correct grid point and travels in the right physical direction in the next time step.

To achieve this there are some corner cases that need to be taken into account explicitly to avoid incorrect reflections, furthermore we make sure that there are no particles residing inside the object at the end of a time step. 

\subsection{Specular reflection steps - requirements and possible breakdown cases}
In this section, we translate the general procedure of the reflection step into a concrete quantum algorithm. Strategies to mitigate the erroneous behavior that might occur for the different corner cases will be discussed on Section \ref{ssec:Correct Reflection}

First we need to identify particles that have reached a grid point inside the obstacle, to ensure that they are placed back into the fluid domain before the start of the next streaming step.
This means that controlled on the current location of the particles (namely inside the obstacle), we set them back onto a grid point outside of the obstacle. Performing such an operation on a quantum computer is, however, far from being trivial since we cannot alter the state of certain qubits controlled on their own states, as this would constitute a non-unitary operation. 
In our case this means that we cannot simply alter the position of the particles in space based on their current position, as that would be precisely attempting an operation on some qubits controlled by themselves. 

A way to overcome this is to use a system of ancillae to facilitate this back placement. This, however, introduces a new complexity since the ancillae would also have to be reset after the positional move has been performed in order to be useable in the next time step.
The resetting of the ancillae to their original position is nontrivial because we clearly cannot use the original requirements to simply flip the ancillae back, as the original control states were positional and we used the ancillae to control a positional change. Instead, we will have to reset the ancillae based on the new location of the particles in combination with their direction and velocities, i.e., we reverse engineer the particles' previous position. 

Another nontrivial aspect of the specular reflection step is to implement it such that all of the reflections are physically correct and we do not encounter unphysical behavior around the corner points. 
If one were to simply reflect the velocity in the $y$-direction upon hitting a $y$-wall and the velocity in the $x$-direction upon hitting an $x$-wall, incorrect behavior around the corner points will emerge. This is because corner points are points inside the object that are part of walls in multiple directions, i.e., in the two-dimensional case they are part of both an $x$-wall and a $y$-wall. 
What will happen in a non-fail-safe implementation of the specular reflection step is that particles hitting such a grid point coming through just the $x$- or $y$-wall have their velocities reflected in both directions, instead of only the one orthogonal to the wall they came through. 

\begin{figure}
    \centering
    \includegraphics[]{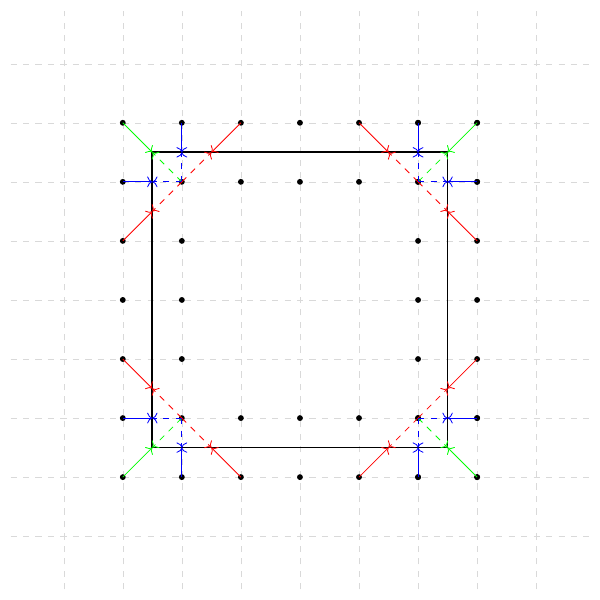}

    \caption{Illustration of the different cases of specular reflection at the corner points of an obstacle. While particles traveling into the obstacle along the blue and green velocity trajectories are reflected correctly even by non-fail-safe reflection algorithms, particles approaching the corner points along the red-arrow trajectories require special treatment as is done in our fail-safe specular reflection algorithm.
 }
    
    \label{fig:example_wrong}
\end{figure}

Figure \ref{fig:example_wrong} gives an example of what can go wrong in a non-fail-safe implementation of the specular reflection step.
We see that in the cases of the green arrows the behavior is correct, as the velocity is reflected in both the $x$- and $y$-direction since the particle hits both an $x$- and $y$-wall. For the case of the blue arrows we see that the behavior is also correct, since the velocity parallel to the wall is zero in this case, therefore reflecting the velocity in this direction has no effect. However, for any particle approaching the corner point from another direction, such as the red arrow, the reflection based on a non-fail-safe implementation will certainly be wrong.  This is due to the fact that such a particle hits a point associated to both an $x$-wall \emph{and} a $y$-wall, even though physically it can easily be seen that the particle only hits either an $x$- \emph{or} a $y$-wall. In a non-fail-safe reflection operation this means that the velocity of the particle is erroneously reversed in both the $x$- \emph{and} the $y$-direction. 

\begin{figure}
    \centering
    \includegraphics[]{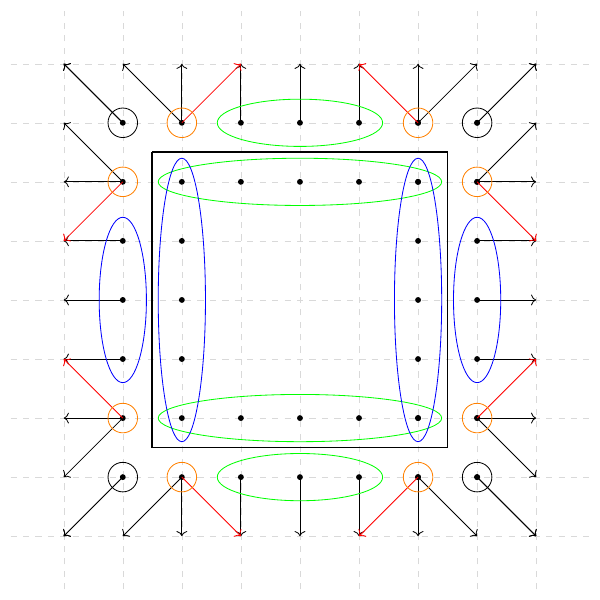}

    \caption{Illustration of all possible corner cases to be taken into account when particles collide with an obstacle (black box) and the physically correct reflection behavior as enforced by our fail-safe specular reflection algorithm.}
    \label{fig:example_right}
\end{figure}

\subsection{Fail-safe specular reflection - 2D case}\label{ssec:Correct Reflection}
In what follows we propose a novel fail-safe specular reflection approach. 
Figure \ref{fig:example_right} shows an obstacle placed on a grid, with both the grid points inside the object and the grid points outside the object drawn. Furthermore, Figure \ref{fig:example_right} shows the possible reflections that can occur around the obstacle. Using this figure we describe how we can implement a unitary operation that treats all possible reflection cases correctly. 

First of all we require $d$ extra ancillae $a_{o,i}$, where $i$ represents the spatial dimension, that facilitate moving the particles out of the obstacle for the next time step and keeping track of which velocity components should be reflected when a particle hits a corner point. Lastly, we make use of the earlier defined ancillae $a_{v,i}$ to correctly set and reset the $a_{o,i}$ ancillae at the beginning and end of the specular reflection step, respectively. Note that using the $a_{v,i}$ ancillae does not add to the complexity of the circuit since we can reuse their state from the streaming step, and we reset them after the specular reflection step instead of directly performing the reset after the streaming operation.

Upon reaching a blue (green) encircled grid point inside the object, the particle has hit an $x$-wall ($y$-wall).
When hitting a grid point that is part of the $x$-wall ($y$-wall) in the object, we flip the ancilla qubit $a_{o,x}$ ($a_{o,y}$) controlled on the $v_{\text{dir},x}$ and $a_{v,x}$ ($v_{\text{dir},y}$ and $a_{v,y}$) qubits. 

Specifically, we flip the extra-defined ancillae only when we just took a step in the direction that the wall reflects, and when we travel in a direction that the wall would reflect based on the position of the object (meaning that if we travel in the negative $y$-direction and we hit a grid point in a $y$-wall in the bottom of the object we do not flip the $a_{o,y}$ ancilla).
This is followed by flipping $v_{\text{dir},x}$ ($v_{\text{dir},y}$) controlled by the $a_{o,x}$ ($a_{o,y}$) ancillae and incrementing the position by one index in the $x$ ($y$) direction. Here, the term incrementing amounts to streaming in the direction $v_{\text{dir},x}$ ($v_{\text{dir},y}$).

The final step consists of resetting the ancillae $a_{o,i}$. The grid points directly outside the object that are not `in the vicinity of a corner point' of the object constitute the trivial case. By `in the vicinity of a corner point' we mean that the current grid point the particle is in, has as adjacent grid point inside the object a grid point that is part of both an $x$-wall and a $y$-wall. In Figure \ref{fig:example_right} the grid points that are not considered to be `in the vicinity of a corner point' are identified as the blue (green) encircled grid points outside of the object. All that needs to be done for them is to reset the $a_{o_x}$ ($a_{o,y}$) ancilla controlled on the position, the direction of the $x$ ($y$) velocity and $a_{v,x}$ ($a_{v,y}$). More specifically, we reset the ancilla for a particle in a grid point that is adjacent to an $x$-wall ($y$-wall) if and only if $a_{v,x}=1$ ($a_{v,y}=1$) and $v_{\text{dir},x}$ ($v_{\text{dir},y}$) points away from the wall. 

Around the corner points of the object we need to be more careful, however. This is due to the fact that the particles could have gotten there via multiple directions, in which case different ancillae would need to be reset. 

For the black encircled grid points we have that both the ancillae $a_{o,x}$ and $a_{o,y}$ need to be reset controlled on $v_{\text{dir},x}$, $v_{\text{dir},y}$, $a_{v,x}$ and $a_{v,y}$. Only when $a_{v,x} = a_{v,y} = 1$ holds and $v_{\text{dir},x}$ and $v_{\text{dir},y}$ are such that they both point away from the object, do $a_{o,x}$ and $a_{o,y}$ need to be reset.

For the orange encirled points we do the following. We first reset the ancillae based on the same criteria as the blue (green) encircled qubits. Subsequently, we note that the respective ancilla should not have been reset only in the case of the red arrow. Therefore, we simply re-reset the ancillae based on the criteria that reflect the case of the red arrows, consisting of position in combination with direction in both the $x$ and $y$ direction ($v_{\text{dir},x}$, $v_{\text{dir},y}$) and whether a step was taken in both directions in the former time step ($a_{v,x}$ and $a_{v,y}$). Specifically, for the arrow located on the highest row on the left, this would mean resetting the ancilla controlled on $v_{\text{dir},x}$ being positive, $v_{\text{dir},y}$ being positive, the position being the dot on the highest row second from the left and $a_{v,x}$ and $a_{v,y}$ both being activated.
This strategy for defining hand-crafted resetting patterns can easily be adapted to the remaining three corners.

\subsection{Fail-safe specular reflection - 3D case}\label{ssec:spec ref 3d}
We will now provide a generalization of our two-dimensional method to three dimensions. The idea of the set-up is the same, we define grid points just inside the object and grid points outside and directly adjacent to the object. 
Then, each grid point inside the object is associated with an $x$-wall or an $y$-wall or a $z$-wall.
Now, if a particle reaches a grid point associated to an $x$-wall ($y$-wall, $z$-wall), was traveling in a direction that this wall would reflect (meaning $v_{\text{dir},x}$ ($v_{\text{dir},y}$, $v_{\text{dir},z}$) was in the right state) and $a_{v,x}=1$ ($a_{v,y}=1$, $a_{v,z}=1$), the ancilla $a_{o,x}$ ($a_{o,y}$, $a_{o,z}$) will be flipped. Notice that here we are using the exact same logic as in the two-dimensional case.

Subsequently the particles are placed back to the adjacent points inside the fluid domain based on the states of $a_{o,x}$, $a_{o,y}$ and $a_{o,z}$. This step is realized in the same way as in the two-dimensional case.

Finally, we need to reset the ancillae $a_{o,x}$, $a_{o,y}$ and $a_{o,z}$. This is again performed by considering the current position of the particle, in combination with their direction and whether or not they were streamed in the last time step. Compared to the two-dimensional case, in three dimensions there are more distinct cases to be taken into account. Instead of distinguishing only between corner points and non-corner points, we need to distinguish between corner points (where three walls come together), points along the edges of the obstacle (where two walls come together) and points on the side of the objects. 

For the points on the side of the object and the points along the edges of the obstacle, the rules for reflection will be the same as in the two-dimensional case. Here the points on the edges of the obstacle behave the same as the corner points in the two-dimensional case, and the points on the side of the objects behave the same as the non-corner points in the two-dimensional case. 

\begin{figure}
\centering
\includegraphics[]{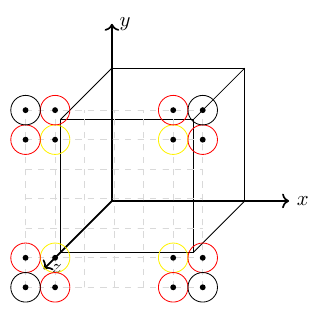}
\caption{Illustration of the corner cases specific to a three-dimensional obstacle. The action applied to the different color-coded grid points is described in Section \ref{ssec:spec ref 3d}.}\label{fig:3dpict}
\end{figure}

\begin{figure}
    \centering
    \includegraphics[]{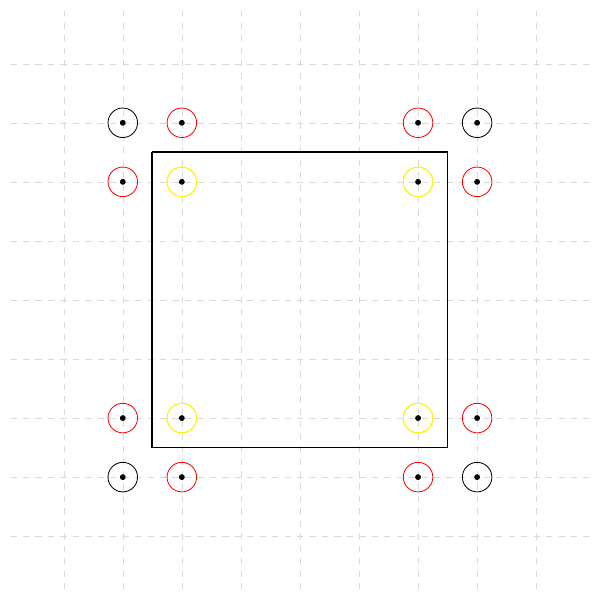}
    \caption{View from the positive $z$-axis onto the $z$-wall of the obstacle. Note that the $z$-wall lies half a grid point in the z-direction below the grid with the encircled points. 
 }\label{fig:3dpictin2d}
\end{figure}

The corner points where three walls come together require a more in depth consideration. Figures \ref{fig:3dpict} and \ref{fig:3dpictin2d} show an object in our three-dimensional grid and uses colors to indicate which grid points need to be taken into account as special cases around corner points.
In the black encircled points the $a_{o,x}$, $a_{o,y}$ and $a_{o,z}$ qubits are reset if the velocity in all three dimensions travels away from the object and the ancillae $a_{v,x}$, $a_{v,y}$, $a_{v,z}$ are all equal to 1. 
In the red encircled points the ancillae of the two dimensions which get reflected by the two walls that intersect are reset, if the particle is moving away from the object in the two respective dimensions and the particle took a step in both dimensions and just before did not travel a step in the third dimension towards the object.
In the yellow encircled point we consider the following behaviors. Here we only reset the ancilla associated with the dimension the wall reflects in if we just took a step in that dimension and the directional qubit points away from the object. Furthermore, we need to check that in none of the other two dimensions we just took a step towards the object. 

\subsection{Efficient object encoding}

In this section we propose a novel approach based on quantum comparison operations for encoding objects in such a way that we can efficiently identify whether or not a grid point is part of a certain object wall. Our approach requires $2(d-1)$ extra ancillae and will be explained at the hand of a two-dimensional example.

Assume that we want to encode an $x$-wall (note that everything works the same to encode an $y$-wall), for example the left-most $x$-wall of Figure \ref{fig:example_right}. Let the grid points just inside the object next to the $x$-wall, i.e., the left-most blue encircled grid points inside the object of Figure \ref{fig:example_right}, range from $[l,u ]$ in the $y$-axis. We first need to determine whether or not we are in one of the left-most blue encircled grid points inside the object with the aid of two quantum comparison operations. The first quantum comparison operation checks whether $y \geq l$ holds. If so we flip the first extra ancilla from its start state $a_{l,1}= 0$ to $a_{l,1}= 1$. Then we check if $y \leq u$ holds, and if so we flip the second extra ancilla to $a_{u,1}= 1$. Now, controlled on these both ancillae and the state of the qubits encoding the position in the $x$ dimension of these left-most blue encircled qubits, we flip the ancilla $a_{o,x}$ indicating that we are in a wall which reflects the velocity in the $x$-direction.
Now we simply reset the $a_{l,1}$ and $a_{u,1}$ ancillae so that they can be reused for other walls, by performing the same quantum comparison operations as described before\footnote{Note that in the example of Figure \ref{fig:example_right} it would be most efficient to first flip the ancilla $a_{o,x}$ as well, controlled on the $a_{l,1}$ and $a_{u,1}$ qubits in combination with the qubits encoding the position in the $x$ dimension being in the position of the right-most blue encircled grid points inside the object. But for now we were only considering the left-most $x$-wall.}.
In the two-dimensional case we perform this operation for all walls in order to set the $a_{o,x}$ and $a_{o,y}$ qubits as required. After completion of this step we continue in the same fashion as described in Section \ref{ssec:Correct Reflection}.

What remains is to reset the $a_{o,x}$ and $a_{o,y}$ qubits after the reflection and streaming steps have been performed which we accomplish, again, with the aid of the quantum comparison operation. To continue with our example of the left-most $x$-wall of Figure \ref{fig:example_right}, we again perform two comparison operations to check whether we are in the left-most blue encircled grid points right outside the wall, which amounts to checking $y \in [l+1,u-1]$. This can be achieved with the same logic as before namely by checking $y \geq l+1$ and $y \leq u-1$. All the other steps and the resetting are performed in the same manner as explained before. The extension to the three-dimensional case is straightforward.

\subsection{Quantum Comparison Operation}\label{ssec: quantum comparison}
In our implementation we use the quantum comparison operation from \cite{Gidney2017}, which compares the integer value $i$ of the $n$-qubit quantum state $\ket{i}$ with a pre-determined constant $k$ saving the result of the comparison in a separate qubit. 

The quantum comparison algorithm works as follows. Say we wish to determine whether the integer value $i$ encoded in basis state $\ket{i}$ is smaller than $k$\footnote{Note that $k \leq 2^n-1$ always holds, since otherwise we trivially know that $k$ is larger than any value that $n$ qubits can encode.}. Then, we first subtract the integer $k$ from the value encoded by the $n+1$ qubits encoding the state $\ket{0}\otimes\ket{i}$, where the prepended $\ket{0}$ qubit will hold the result of the computation. Now there are two cases, $i\geq k$ and $i<k$. \\

Assume that we are in the case $i \geq k$. Then subtracting $k$ from $i$ encoded in  $\ket{0}\otimes\ket{i}$ gives $\ket{0}\otimes \ket{i-k}$, which leaves the prepended $\ket{0}$ qubit unchanged. 
Now we simply perform an addition of the value $k$ to the $n$-qubits that were used to encode $i$ and now encode $\ket{i-k}$. This gives $\ket{0}\otimes\ket{i-k+k} = \ket{0}\otimes\ket{i}$. And so in total we end up with a quantum register in the state $\ket{0}\otimes\ket{i}$. Where the $\ket{0}$ state is the result of the computation which means that $i \geq k$ in fact holds, and the last $n$ qubits are again in the original state $\ket{i}$. \\

Now let us assume that we are in the second case $i<k$. Again, we start by periodically subtracting the integer $k$ from $i$ encoded in the state $\ket{0}\otimes\ket{i}$, only now since $k>i$ we end up flipping the state of the prepended $\ket{0}$ qubit. This is because periodically subtracting $k$ from $i$ in a state encoded by $n+1$ qubits results in the state $\ket{2^{n+1} -1 - k +i}$. Since $k \leq 2^n-1$ and $i \geq 0$ we must have that $2^{n+1} -1 - k +i \geq 2^n$ and so the state of the most significant qubit must be equal to $\ket{1}$. This means that the total qubit register is now in the state $\ket{1} \otimes \ket{2^{n+1} -1 - k +i - 2^n} =\ket{1} \otimes \ket{2^{n} -1 - k +i }  $. 

Then, as before, we simply add the integer $k$ again to the last $n$ qubits of the register leaving us with $\ket{1} \otimes \ket{2^{n} -1 +i } = \ket{1} \otimes \ket{ i } $ due to periodicity. Notice how the most significant qubit is left in the state $\ket{1}$ indicating that in fact $k > i$, whilst again the qubits encoding $\ket{i}$ have not changed.  

Multiple realizations of quantum addition and subtraction operations are described in the literature \cite{Häner2017,Takahashi2009,Cuccaro2004,Draper1998} and can be used for the comparison operation, all having their own complexities and required ancillae associated to them. \\

A quantum primitive for checking whether $i\geq k$ holds can be easily designed from the above ($i<k$) by simply negating the qubit that holds the result after performing $i<k$. 

Implementing a quantum primitive that determines $i\leq k$ is a bit more involved as we need to consider two different cases. First, if $k = 2^n -1$ simply flip the ancilla holding the result, as $i \leq k$ trivially holds, otherwise we can implement $i< k+1$ to get the desired result. Notice that here we can easily implement this two case method for $i \leq k$ since $k$ is an integer determined before the running of the algorithm.   

\section{Results} \label{sec:results}
To demonstrate the correct functioning of our \textcolor{black}{QTM} we implemented the proposed algorithm in Qiskit \cite{Qiskit} and performed some preliminary simulation runs for a moderately small grid in two spatial dimensions. All runs were performed with Qiskit's local quantum simulator Aer on an 8-core Intel i7-10610U CPU running at 1.8GHz. For simplicity, we assumed perfect qubits, i.e., no noise model and all-to-all qubit connections, i.e., a generic QPU.

Our overall algorithm consumes just 22 qubits in total, 6+6=12 qubits for the grid locations, 2+2=4 qubits to encode the velocity vectors ($N_v=2$ in both directions), and 6 ancillae. ased on the amount of qubits, today's quantum computers should be able to execute our \textcolor{black}{QTM} solver, however, the circuit depth exceeds capacities of todays devices by orders of magnitude so that we are only able to show results produced on a quantum computer simulator.

\begin{figure}[ht]
\begin{subfigure}{.5\textwidth}
  \centering
\includegraphics[width=\textwidth]{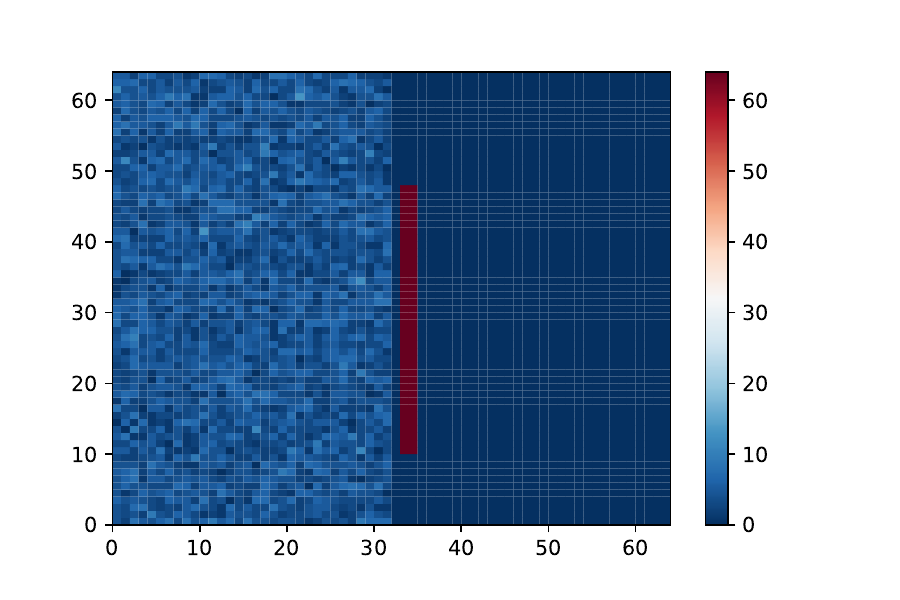}
\caption{Output after 0 timesteps.}\label{fig:plaatje 0}
\end{subfigure}
\begin{subfigure}{.5\textwidth}
  \centering
\includegraphics[width=\textwidth]{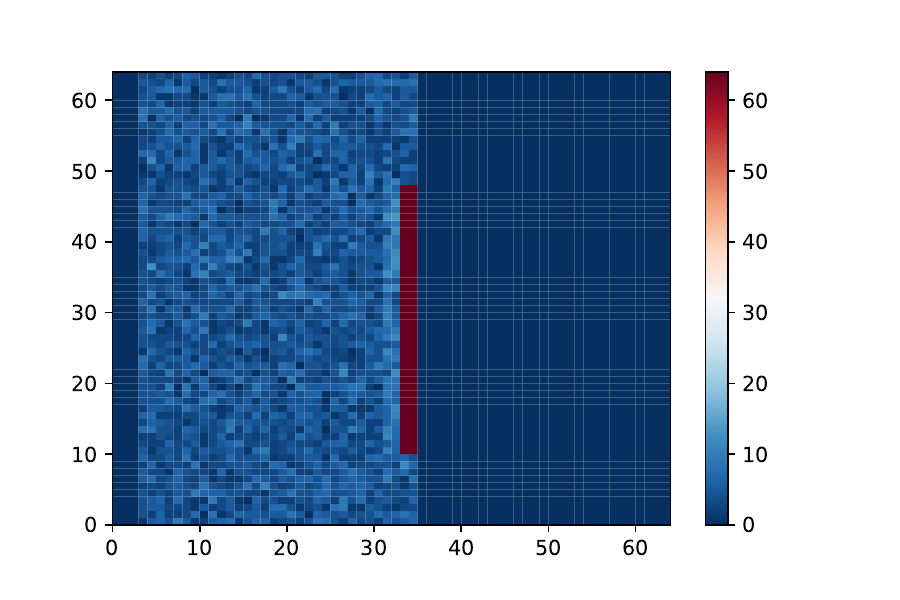}
\caption{Output after 3 timesteps.}\label{fig:plaatje 3}
\end{subfigure}

\begin{subfigure}{.5\textwidth}
  \centering
\includegraphics[width=\textwidth]{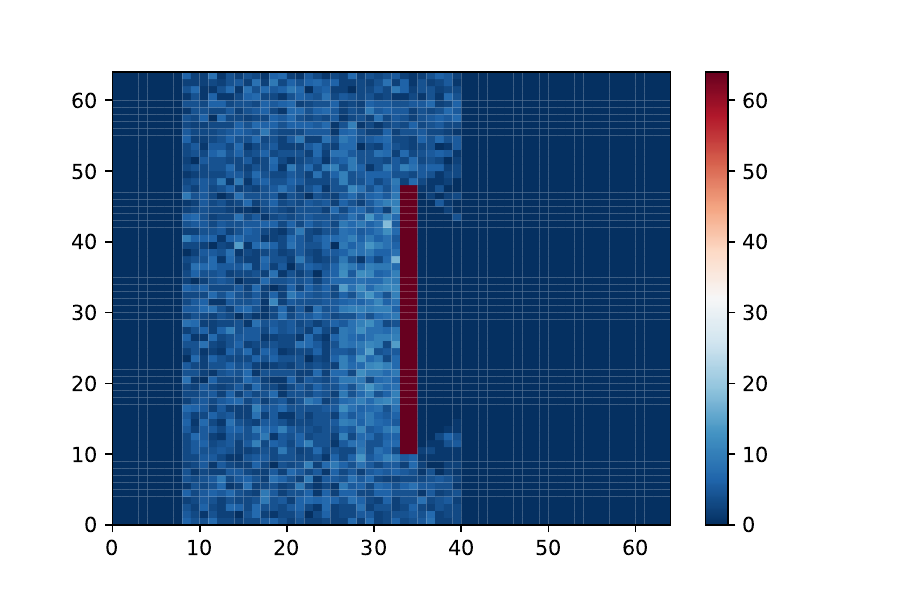}
\caption{Output after 8 timesteps.}\label{fig:plaatje 8}
\end{subfigure}
\begin{subfigure}{.5\textwidth}
  \centering
\includegraphics[width=\textwidth]{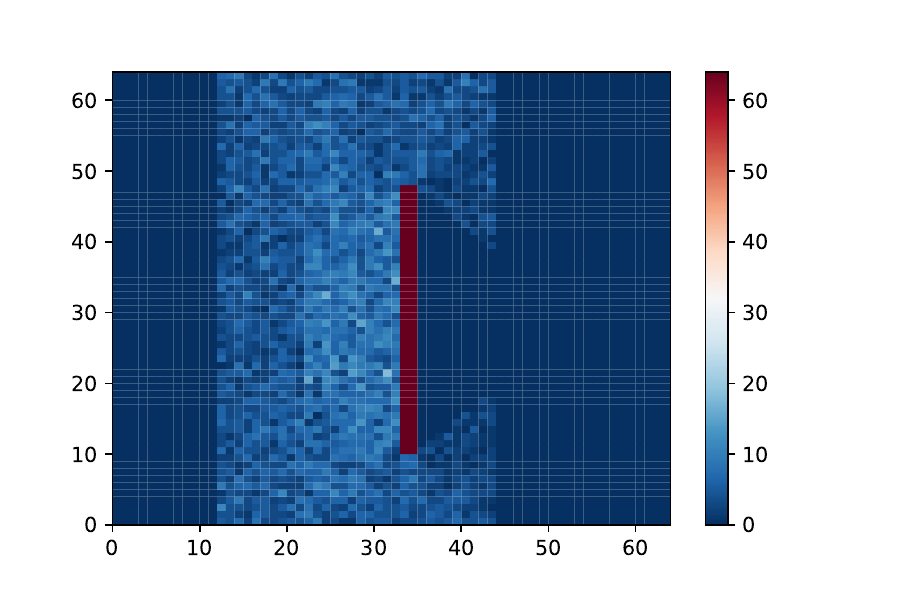}
\caption{Output after 12 timesteps.}\label{fig:plaatje 12}
\end{subfigure}

\begin{subfigure}{.5\textwidth}
  \centering
\includegraphics[width=\textwidth]{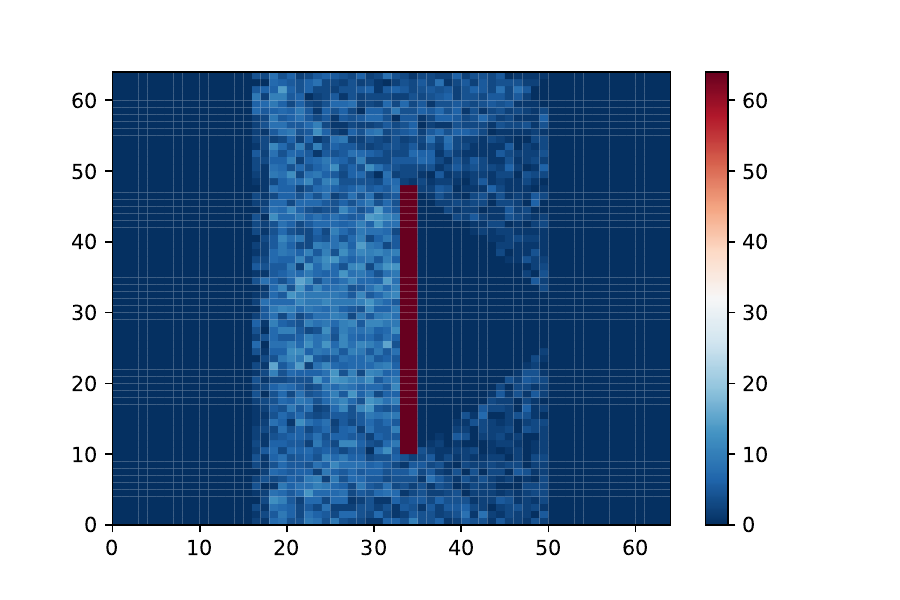}
\caption{Output after 18 timesteps.}\label{fig:plaatje 18}
\end{subfigure}
\begin{subfigure}{.5\textwidth}
  \centering
\includegraphics[width=\textwidth]{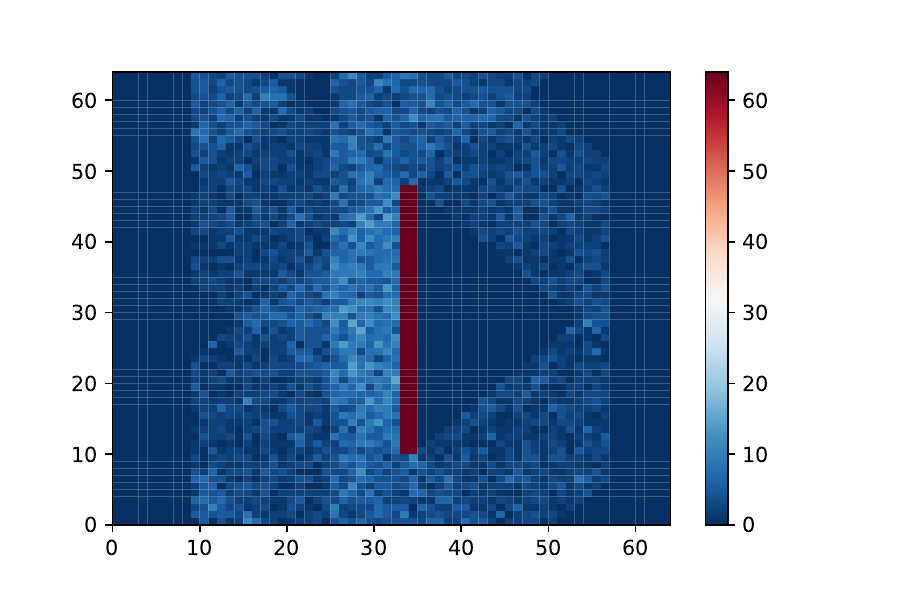}
\caption{Output after 25 timesteps.}\label{fig:plaatje 25}
\end{subfigure}
\caption{Sequence of solutions computed on a $64\times 64$ grid by the proposed \textcolor{black}{QTM} solver on 22 simulated qubits adopting 8.192 measurement.}
\label{fig:qcfd_results_8192}
\end{figure}

Figure \ref{fig:qcfd_results_8192} shows a sequence of numerical results computed on a $64\times 64$ grid with an internal object of size $3\times 39$ located with its lower left corner at position $(34,11)$. The initial state was prepared by applying Hadamard gates to all the qubits encoding the grid in the $y$-dimension and all but the most significant qubit encoding the grid in the $x$-dimension, that is, all particles are equidistributed in the left half of the fluid domain, whereas the right half is in vacuum state. Perfectly reflecting boundary conditions are prescribed at the internal obstacle while periodic boundary conditions are prescribed at all four domain boundaries.

In the preparation step of the quantum state we also applied a Hadamard gate the the $v_{\text{dir},y}$ ancilla, spreading out the direction of the velocities of the particles in the $y$-dimension. Finally, we apply a NOT gate to the $v_{\text{dir},x}$ ancilla, leading to the particles traveling in the positive $x$-direction. Figures \ref{fig:plaatje 3}--\ref{fig:plaatje 25} illustrate how particles move in the positive $x$-direction and in both the positive and negative $y$-directions filling the vacuum behind the obstacle as expected.

In order to stay as close as possible to the capabilities of a physical quantum computer we performed 8.192 shots. Knowing the exact size and position of the obstacle we excluded 117 from the 4.096 possible states from the measurement so that, on average, each grid point gets measures twice.

\begin{figure}[ht]
\begin{subfigure}{.5\textwidth}
  \centering
\includegraphics[width=\textwidth]{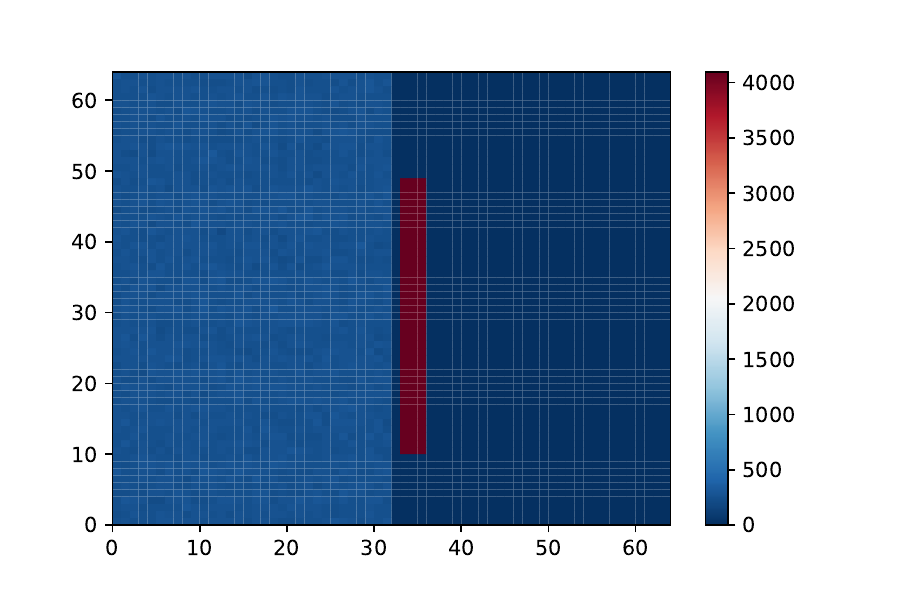}
\caption{Output after 0 timesteps.}\label{fig:plaatje mm 0}
\end{subfigure}
\begin{subfigure}{.5\textwidth}
  \centering
\includegraphics[width=\textwidth]{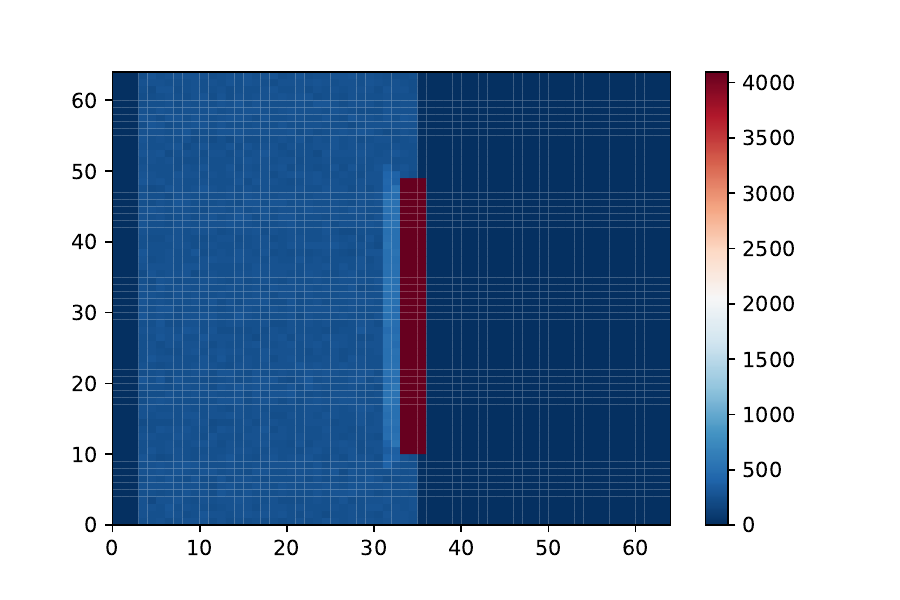}
\caption{Output after 3 timesteps.}\label{fig:plaatje mm 3}
\end{subfigure}

\begin{subfigure}{.5\textwidth}
  \centering
\includegraphics[width=\textwidth]{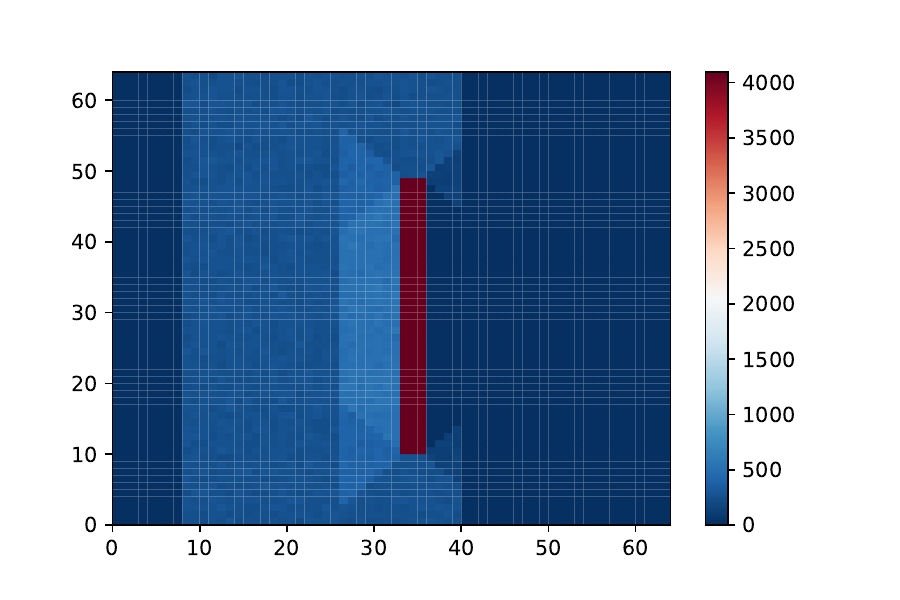}
\caption{Output after 8 timesteps.}\label{fig:plaatje mm 8}
\end{subfigure}
\begin{subfigure}{.5\textwidth}
  \centering
\includegraphics[width=\textwidth]{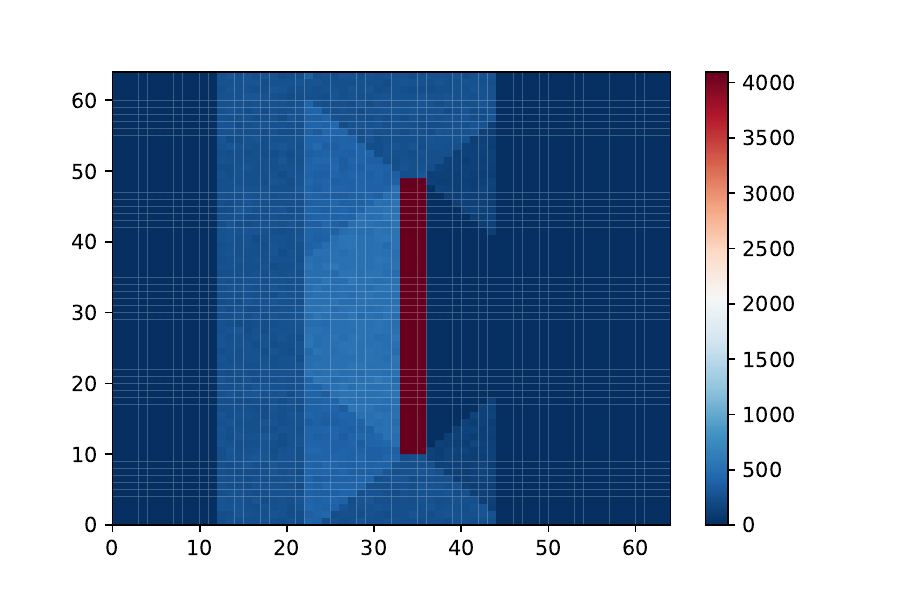}
\caption{Output after 12 timesteps.}\label{fig:plaatje mm 12}
\end{subfigure}

\begin{subfigure}{.5\textwidth}
  \centering
\includegraphics[width=\textwidth]{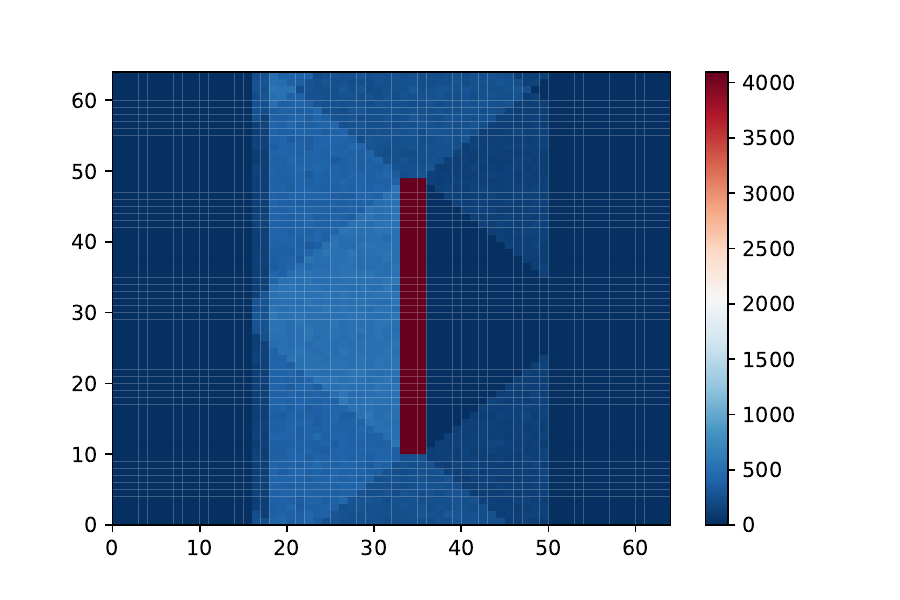}
\caption{Output after 18 timesteps.}\label{fig:plaatje mm 18}
\end{subfigure}
\begin{subfigure}{.5\textwidth}
  \centering
\includegraphics[width=\textwidth]{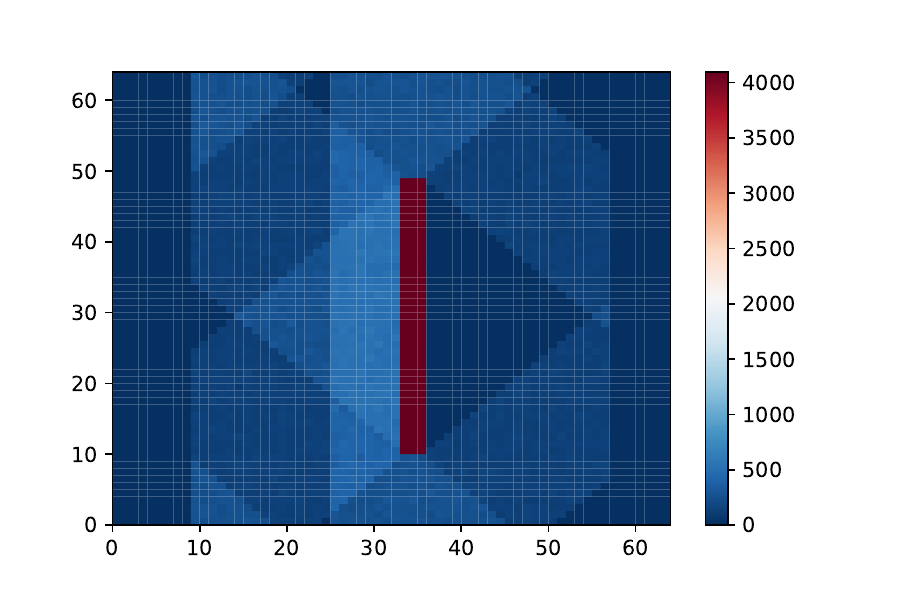}
\caption{Output after 25 timesteps.}\label{fig:plaatje mm 25}
\end{subfigure}
\caption{Sequence of solutions computed on a $64\times 64$ grid by the proposed \textcolor{black}{QTM} solver on 22 simulated qubits adopting 524.288 measurements.}
\label{fig:qcfd_results_524288}
\end{figure}

Figure~\ref{fig:qcfd_results_524288} depicts the same sequence of results with 524.288 measurements showing a much better enunciation of the flow pattern. All plots show the density of particles at the respective grid points. Future research will focus on problem-specific measurements of application specific integral quantities of interest that might require much less measurements.

\section{Complexity analysis}\label{sec:comparison to state of the art}
In this section we give a detailed complexity analysis of our approach. 
Since most quantum hardware can only implement single- and two-qubit gates natively \cite{Rigetti} \cite{IBM}, we decompose the multi-qubit operations into two-qubit gates to find a realistic quantum complexity for the proposed method. Specifically we choose to give the complexity in terms of CNOT gates, while CNOT is one of the native gates for IBM QPUs \cite{IBM} it needs to be emulated by other two-qubit gates on other quantum hardware such as the QPUs from Rigetti \cite{Rigetti}. However, in these cases the CNOT gate can be decomposed using a fixed number of the natively implemented two-qubit gates, so this does not effect the overall complexity. 

We subsequently compare our found complexity to the complexity of the current best known quantum algorithm for the \textcolor{black}{transport} equation \cite{Todorova2020}. In doing so we show that our method not only outperforms the current state-of-the-art by implementing the quantum reflection step in a fail-safe manner, but we also reach a better complexity.

\subsection{Complexity of multi-controlled NOT operations}\label{ssec:multinot compl}
In order to provide a cost analysis for our algorithm, we first determine the cost associated to implementing a multi-controlled NOT gate, denoted as C$^p$NOT in the following, in terms of CNOT gates. 
As the decomposition of C$^p$NOT gates into CNOT gates is a field of active research, we selected a few representative approaches, namely, one without ancilla qubits, one with a linear amount of ancilla qubits and multiple with a constant amount of ancilla qubits. \\

A decomposition without any ancilla qubits was given by Barenco et al. in \cite{Barenco1995}. The reported decomposition requires $48(p+1)^2 + \Theta(p)$ `basic operations'\footnote{Here the term basic operations implies either a CNOT or single qubit gate.}, from which we deduce that approximately $c p^2$ with $22<c<48$ CNOT gates will be required leading to a complexity of $\mathcal{O}\left (p^2\right )$ CNOT gates. \\

Nielsen and Chuang present an alternative decomposition with a linear amount of ancilla qubit \cite{mikeandike}. The decomposition of a single C$^p$NOT gate requires $p-1$ ancillae and $2(p-1)$ Toffoli gates\footnote{A Toffoli gate is a CCNOT gate, i.e., a controlled NOT gate with two control qubits.}. Each Toffoli gate in turn is decomposed using 6 CNOT gates (and some other operations) \cite{mikeandike}. Therefore, a total of $12(p-1)$ CNOT gates is required to decompose a C$^p$NOT operation into native one- and two-qubit gates. \\

As a third alternative the paper by Barenco et al. \cite{Barenco1995} provides a decomposition of C$^p$NOT operations for $p\geq 5$ using a single ancilla, $8(p-3)$ Toffoli gates and $ 48 (p+2)$ `basic operations' leading to $c p$ CNOT gates with $70 \leq c \leq 96$.

In addition to the three approaches mentioned above, we can also use the quantum comparison operation discussed in Section \ref{ssec: quantum comparison} in combination with only one ancilla qubit. In fact, one can find whether or not the $p$ control qubits are in the state $\ket{q_1 \dots q_p} = \ket{1 \dots 1}$ by using the comparison operation to check whether $q \geq 2^p-1$ holds, where $q$ is the integer value encoded by the qubits $q_1 \dots q_p$. The cost of such a comparison operation depends on the costs of the constant addition operation. Using the Draper adder as described in Section \ref{sec:efficient quantum convection operation} this method requires $4p^2 + 6p + 3$ CNOT operations\footnote{Since a QFT operation applied to $k$ qubits consists of the following two-qubit gates, first $\frac{k(k-1)}{2}$ controlled phase shift gates followed by $\lfloor k/2 \rfloor$ swap gates. A swap gate can be decomposed using 3 CNOT operations and a controlled phase shift operation requires two CNOT operations. Therefore the total cost of a QFT operation applied to $k$ qubits is $k(k-1) + \lfloor \frac{3k}{2} \rfloor$ CNOT operations. Then the total costs given in the text follows from the fact that the Draper Adder implemented comparison operation consists of first a QFT and QFT$^\dagger$ applied to $p+1$ qubits followed by  a QFT and QFT$^\dagger$ applied to $p$ qubits.}. Therefore this method requires $\mathcal{O} \left ( p^2\right )$ CNOT gates and one additional ancilla.

This implementation appears to match the complexity of the `recursion method', provided by Qiskit \cite{ Qiskit}, which also requires 1 ancilla qubit. 
IBM does not, however, give a complexity analysis in terms of the amount of CNOT gates required or a reference detailing the implementation. We numerically analyzed the complexity of this implementation by running it for several values of $p$ (up to $p=40)$ and found experimentally that it requires $\approx 2 p^2$ CNOT gates. \\

Having, in our opinion, the best trade-off between the amount of CNOT gates and ancilla qubits required for realistic values of $p$, we choose to use the `comparison' or `recursion' multi-controlled NOT decomposition in our further analysis. 

We would like to stress here that there might be cheaper methods for implementing a multi-controlled NOT operation possible as this is a topic of active research. This will only decrease the total cost of our method and it will not change the conclusion presented in Section \ref{ssec:comp_soa_q} of our method having the lowest complexity of any known \textcolor{black}{QTM}. 

\subsection{Complexity of our \textcolor{black}{QTM} solver}\label{ssec:multi not impl}
In this section we give a detailed analysis of the costs of our \textcolor{black}{QTM} solver by first analyzing the different steps separately and, subsequently, combining these results to find the total complexity of the algorithm. 

\subsubsection{Complexity of quantum state preparation}\label{sssec:stateprep}
In order to run the \textcolor{black}{QTM} solver, we first need to prepare the input state to represent the desired particle distribution and velocities, at the initial time $t=0$.
As it is known that the preparation of arbitrary quantum states can become exponentially expensive \cite{Plesch2011, mikeandike}, we restrict ourselves for the numerical results presented in this paper to equidistributed particles and mention a few publications that describe efficient preparation procedures for specific quantum state distributions.

Preparing general sparse quantum states is an active area of research with promising results \cite{Zhang2022, deVeras2022,gleinig2021}, since our input quantum state is typically highly sparse, this is potentially useful for the efficient state preparation of our quantum states.
Another interesting result for the input state preparation is given by Grover and Rudolph \cite{Grover2002}, who present an efficient process for preparing quantum states that form a discrete approximation of an efficiently integrable probability density function, such as log-concave distribution functions. Since the Maxwell-Boltzmann distribution is log-concave our future research will focus on exploiting this property for the efficient quantum state preparation.

\subsubsection{Complexity of the streaming step} \label{sssec:compl conv step}
The quantum streaming step consists of a streaming operation applied to the particles traveling at a pre-determined velocity. In order to implement this we first apply a $\text{C}^p\text{NOT}$ gate between the velocity qubits in each dimension $i$ and the $a_{v,i}$ ancilla qubits. Each $\text{C}^p\text{NOT}$ gate has a total of $p=n_{v_i}-1$ control qubits. We implement this operation for the particles whose velocity magnitudes are marked for advancement in that particular timestep. The amount of velocity magnitudes being advanced in each timestep $n^v_t$ can trivially be capped by $N_v$, but in practice will be equal to $1$ in $89.6\%$ of the cases. We found this percentage by performing numerical analysis for $N_v$ up to 1024. 
Subsequently, the increase operation is applied controlled on the $a_{v,i}$ qubits.

Our increase operator consists of a QFT operation applied to the $n_{g_i}$ qubits in each dimension $i$, followed by a phase shift on the $n_{g_i}$ qubits and a final QFT$^\dagger$ operation. In each dimension $i$, a QFT operation requires $\mathcal{O}(n_{g_i}^2)$ two-qubit operations \cite{Musk2020}, the phase shifts applied to the $n_{g_i}$ qubits is controlled on the $a_{v,i}$ ancilla qubits and so this operation costs $\mathcal{O}(n_{g_i}^2)$ two-qubit operations. 
Therefore, our increase operator only requires $\mathcal{O}(n_{g_i}^2)$ CNOT operations, controlled on the $a_{v,i}$ qubits in each dimension $i$, which allows us to cap its complexity by $\mathcal{O}(d n_{g_\text{max}}^2)$ CNOT operations.
In total that leaves us with a complexity of the convection step of $\mathcal{O} \left ( d n^v_t \right )$ in the amount of C$^p$NOT operations with $p=n_{v_\text{max}}-1$ plus $\mathcal{O} \left ( d n_{g_\text{max}}^2 \right )$ CNOT operations.\\

Using the complexity analysis of the multi-controlled NOT operations provided in Subsection \ref{ssec:multinot compl}, we get an overall complexity of $$\mathcal{O} \left ( d n^v_t n_{v_\text{max}}^2 + d n_{g_\text{max}}^2 \right )$$ in the amount of CNOT gates for the convection step.

\subsubsection{Complexity of the specular reflection step}
The specular reflection step for each wall starts with two quantum comparison operations, applied to the qubits encoding the location in a single dimension with the $a_{l}$ and $a_{u}$ qubits storing the result. This leads to a complexity of $\mathcal{O} \left ( n_{g_\text{max}}^2 \right) $ CNOT gates when using the Draper adder \cite{Draper1998} to implement the comparison operations. 

Subsequently a multi-controlled NOT gate applied to the ancilla qubits $a_{o,i}$, $i \in \{1,\dots,d\}$, controlled on the position in the dimension the considered wall reflects, the $2(d-1)$ $a_{u}$, $a_{l}$ ancillae and one $a_{v,i}$ qubits and one directional qubit $v_{\text{dir},i}$ is applied. Therefore the second step of the specular reflection for each wall consists of a $\text{C}^{p}\text{NOT}$ operations with $p \leq n_{g_\text{max}} +2\left ( d-1 \right ) +2 =  n_{g_\text{max}} +2d$. 

Since the two steps above are applied to each of the $n_w$ walls we get a total complexity of $\mathcal{O} \left ( n_w n_{g_\text{max}}^2 \right) $ CNOT gates and $\mathcal{O}\left ( n_w \right )$ $\text{C}^{p}\text{NOT}$ operations with $p \leq n_{g_\text{max}} +2d$.

Then, controlled on the $a_{o,i}$ qubits we perform a NOT operation on the $v_{\text{dir},i}$ qubits. This operation consists of $d$ CNOT gates, and so the complexity is $\mathcal{O}\left ( d\right )$ CNOT gates. This is followed by a single incrementation operation applied to the positional qubits controlled on $a_{o,i}$ in each dimension. 
The incrementation operation again has complexity $\mathcal{O} \left (d n_{g_\text{max}}^2 \right )$ CNOT operations. 

Subsequently, we reset the $a_{o,i}$ ancillae. Which is established for each wall by first performing two quantum comparison operations, followed by a multi-controlled NOT gate targeting the $a_{o,i}$ ancillae controlled on the qubits encoding the position in a single dimension in combination with the $2(d-1)$ $a_{u}$, $a_{l}$ ancillae, one $a_{v,i}$ qubit and one directional qubit $v_{\text{dir},i}$. This leaves us with a total complexity of again $\mathcal{O} \left ( n_w n_{g_\text{max}}^2 \right) $ CNOT gates and $\mathcal{O}\left ( n_w \right )$ $\text{C}^{p}\text{NOT}$ operations with $p \leq n_{g_\text{max}} +2d$.

Only for the resetting of the $a_{o,i}$ qubits we need to take into account the special rules for resetting the corner cases by applying a multi-controlled NOT gate targeting an ancilla $a_{o,i}$, controlled on the position as well as the $a_{v,i}$ qubits and the direction qubits $v_{\text{dir},i}$. Therefore the complexity of resetting an ancilla $a_{o,i}$ for a corner case becomes $\mathcal{O}\left (  1 \right )$ $\text{C}^{p}\text{NOT}$ operations with $p \leq n_g +2d$. Since the amount of corner cases is linear in the amount of walls the total complexity becomes $\mathcal{O}\left (  n_w  \right )$ $\text{C}^{p}\text{NOT}$
operations with $p \leq n_g +2d$.

At the end of the specular reflection step, we reset the $a_{v,i}$ ancilla qubits in each dimension $i$, using a $\text{C}^p\text{NOT}$ gate with $p=n_{v_i}-1$. 
As explained in Section \ref{sssec:compl conv step} the total costs of this operation amount to $\mathcal{O} \left ( d n^v_t \right )$ C$^p$NOT operations with $p=n_{v_\text{max}}-1$. \\

In total, the complexity of the reflection step is equal to
$\mathcal{O} \left ( n_w n_{g_\text{max}}^2 \right) $ CNOT gates plus $\mathcal{O}\left ( n_w \right )$ $\text{C}^{p}\text{NOT}$ operations with $p \leq n_{g_\text{max}} +2d $ plus $\mathcal{O}\left (  n_w  \right )$ $\text{C}^{p}\text{NOT}$
operations with $p \leq n_g +2d$ plus $\mathcal{O}\left ( d\right )$ CNOT gates plus $\mathcal{O} \left (d n_{g_\text{max}}^2 \right )$ CNOT gates. 

When combining these we get a total complexity of $\mathcal{O} \left ( (n_w + d)  n_{g_\text{max}}^2  \right)  $ CNOT gates
plus $\mathcal{O}\left (  n_w  \right )$ $\text{C}^{p}\text{NOT}$
operations with $p \leq n_g +2d$, for the complexity of the reflection step. 

Translating these results in terms of multi-controlled NOT complexities into single-controlled NOT complexities using the results from Subsection \ref{ssec:multinot compl}, we get the following CNOT gate complexity 
$$\mathcal{O}\left (  n_w \left ( n_g +2d \right)^2 + (n_w + d)  n_{g_\text{max}}^2 \right  ) = \mathcal{O}\left (  n_w n_g^2+ (n_w + d)  n_{g_\text{max}}^2 \right ) = \mathcal{O}\left (  n_w n_g^2 \right ).$$

\subsubsection{Total \textcolor{black}{QTM} complexity}
Combining the complexities of the convection and specular reflection step as detailed above, we obtain that the total complexity of the algorithm per timestep amounts to $\mathcal{O} \left ( (n_w + d)  n_{g_\text{max}}^2  \right)  $ CNOT gates and $\mathcal{O} \left ( n_w \right )$ $\text{C}^{p}\text{NOT}$ gates with $p \leq n_g +2d$ and $\mathcal{O} \left ( d n^v_t \right )$ in the amount of C$^p$NOT operations with $p=n_{v_\text{max}}-1$.  \\

Using the multi-controlled NOT decomposition as before, this leads to
$$\mathcal{O}\left (  n_w n_g^2 + d n^v_t n_{v_\text{max}}^2 + d n_{g_\text{max}}^2 \right ) =\mathcal{O}\left (  n_w n_g^2 + d n^v_t n_{v_\text{max}}^2 \right )$$ CNOT gates per time step.

\subsection{Complexity of alternative \textcolor{black}{QTM} implementations}\label{ssec:comp_soa_q}
An alternative \textcolor{black}{quantum algorithm for the transport} equation presented in the literature is the one by Todorova and Steijl \cite{Todorova2020}. Though not being fully fail-safe in the specular reflection step, the authors present some complexity analysis which led us to perform a rigorous comparison of the complexities of both approaches.

As reported in their paper, Todorova and Steijl conclude a complexity in terms of $\text{C}^p\text{NOT}$ gates of $\mathcal{O}\left ( d N_v \log_2 \left ( D / h  \right ) \right )$ for the streaming step per timestep with $p=n_{g_\text{max}} + n_{v_\text{max}}$. The complexity of their specular-reflection boundary condition, again, quantified in terms of C$^p$NOT gates, is $\mathcal{O}\left ( d N_v \log_2 \left ( D / h  \right ) \right )$. Here, the authors do not provide an explicit estimation for $p$, but since they control on positions in all $d$ spatial dimensions and streaming speeds combined, we derive a complexity of $p = n_{g} +n_{v_\text{max}}$.

Since assumptions and parameters in the paper by Todorova and Steijl differ from ours, it is not immediately clear how to compare their complexity analysis with ours on a fair basis. 
Rewriting their complexity result in terms of the parameters adopted in our analysis yields $\mathcal{O}\left ( d N_v \log_2 \left ( D / h  \right ) \right ) = \mathcal{O}\left  (d N_v n_{g_\text{max}} \right ) $ C$^p$NOT operations with $p \leq n_{g} +n_{v_\text{max}}$.
Finally, rewriting their complexity result in terms of single-controlled CNOT rather than C$^p$NOT gates, yields a total complexity of $\mathcal{O}\left ( d N_v n_{g_\text{max}} (n_{g} +n_{v_\text{max}})^2 \right )$.

Another difference that makes our complexity hard to compare with the complexity of the algorithm by Todorova and Steijl is that the aforementioned authors do not adopt the variable $n_v^t$, and instead use the maximum value $N_v$. For ease of comparison we will replace $n_v^t$ with $N_v$ in our estimates which is possible as $n_v^t \leq N_v$ holds trivially.
Finally, Todorova and Steijl also do not take the amount of walls into account explicitly, but instead seem to consider it a constant, we need to relax this parameter in our complexity estimation as well. With all the aforementioned changes in place we can rewrite our leading complexity terms as $\mathcal{O}\left ( n_g^2 + d N_v n_{v_\text{max}}^2 \right )$ CNOT operations, which is still significantly cheaper than the \textcolor{black}{QTM} approach by Todorova and Steijl having a total CNOT complexity of $\mathcal{O}\left ( d N_v n_{g_\text{max}} (n_{g} +n_{v_\text{max}})^2 \right )$.

\subsection{Complexity comparison of incrementation operations}\label{ssec:complexity incrementation}
One of the improvements we propose in this paper with respect to other known method \cite{Todorova2020} is the use of the Quantum Draper Adder (QDA) \cite{Draper1998} inspired approach to implement the quantum incrementation operation, leading to a cheaper quantum primitive for the streaming operation. In this section we provide a comparison of the costs in terms of CNOT gates of the incrementation method with our QDA inspired approach compared to the approache implemented in \cite{Todorova2020}. 

Consider the QDA inspired incrementation operation applied to the qubits $g_i$ encoding the position in dimension $i$. It consists of a QFT followed by a phase shift gate and finally a QFT$^\dagger$ operation. 
The cost of implementing the QFT (QFT$^\dagger$) operation is $n_{g_i}^2 + \frac{1}{2}n_{g_i}$ CNOT operations as shown in Section \ref{ssec:multinot compl}. Therefore our incrementation operation can be applied at a total cost of $2 n_{g_i}^2 + n_{g_i}$ CNOT operations. 

The cost of the quantum incrementation operation proposed in \cite{Todorova2020} consists of $\sum_{p=0}^{n_{g_i}-1} \text{C}^p\text{NOT}  $ gates. Only when assuming that a C$^p$NOT can be decomposed by $\mathcal{O}\left ( p\right )$ CNOT gates, we find that the total costs of the incrementation operation implemented in these papers have the same order of magnitude with respect to the amount of CNOT gates as our QDA inspired incrementation method. To the best of our knowledge, however, a decomposition of a C$^p$NOT gate that requires $\mathcal{O}\left ( p\right )$ CNOT gates, either requires an extra $p-1$ ancilla qubits or requires a single ancilla qubit and $c$ CNOT-operations with a large constant $c$, making our method more efficient \ref{ssec:multinot compl}.

\subsection{Tabular overview of complexities}
For the ease of the reader we have provided an overview of the complexities of the different steps of our QTM as well as that of the the other quantum transport method by Todorova and Steijl (T \& S) \cite{Todorova2020}. In order to provide a fair comparison of the different algorithms we have rewritten the complexities given in each paper to use the same variables.

\noindent \resizebox{\linewidth}{!}{%
\begin{tabular}{ | c || c | c | c|} \hline
Method & \makecell{Complexity \\ per timestep} &  Streaming & Reflection  \\
\hline
\hline
    S \& M & $\mathcal{O}\left ( n_g^2 + d N_v n_{v_\text{max}}^2 \right )$  & $\mathcal{O} \left ( d n^v_t n_{v_\text{max}}^2 + d n_{g_\text{max}}^2 \right )$ & $\mathcal{O}\left (  n_w n_g^2 \right )$  \\
    \hline
    T \& S & $\mathcal{O}\left ( d N_v n_{g_\text{max}} (n_{g} +n_{v_\text{max}})^2 \right )$ & $\mathcal{O}\left ( d N_v n_{g_\text{max}} (n_{g} +n_{v_\text{max}})^2 \right )$ & $\mathcal{O}\left ( d N_v n_{g_\text{max}} (n_{g} +n_{v_\text{max}})^2 \right )$  \\
\hline
\end{tabular}}

\section{Conclusion and outlook}

Our detailed complexity analyses show that already a moderate number of a few dozens to a hundred fault-tolerant qubits suffice to solve the transport equation in 2 and 3 dimensions on a quantum computer. As our primary focus lies on implementability on upcoming fault-tolerant quantum computers, all algorithmic steps are optimized towards directly implementable single- and two-qubits gates, instead of theoretically more elegant but hard-to-implement multi-controlled operations that call for decomposition into native gates. Next to that we optimize the encoding of the discrete velocity to allow for a more efficient implementation of the reflection operation, the implementation of which we have made fail-safe to allow for physically correct behavior upon collision with a wall. 
Furthermore we have shown that our approach is significantly more efficient than state of the art quantum methods for the transport equation.

Since we are only at the start of exploring the potential of quantum computers for simulating flow problems, there are still quite some restrictions to the current method such as being limited to particles having relative velocities in each dimension smaller than 1, having a total simulated time dependent on the velocities present and using Cartesian grids where object walls are aligned with the grid. In order for the field to develop and grow future work should tackle these issues and develop algorithms without those restrictions.
Another cornerstone on the way to quantum algorithms for realistic CFD applications is the treatment of nonlinear behavior, as it is often encountered in real world flows. 

Further development of quantum SDK's that allow for a straightforward implementation of the described algorithm will also be required to bring QCFD to its stage of practical implementability and allow for large scale simulations and comparisons to be performed.
The final key element that needs to evolve before QCFD can reach its full potential is the fidelity of quantum hardware, as fault-tolerant computers are required for such methods but not yet available. This, of course, lies beyond the control of the computational scientists and is for the experimentalists to realise.

\section*{Acknowledgment}

The authors would like to thank Dr. David de Laat (TU Delft) for fruitful discussions and valuable feedback on the manuscript.

\printbibliography
\newpage
\appendix
\section{Quantum Computing}\label{app:quantum computing}
In this appendix we introduce the fundamentals of quantum computing necessary to quantum fluid dynamics to the non-expert. This is a brief introduction, intended to supply the minimum amount of information and understanding required for the comprehension of our paper. For further studying we suggest using the book by Nielsen and Chuang \cite{mikeandike} and the lecture notes by De Wolf \cite{DeWolf}.

\subsection{Qubits}
Quantum computers are built up using the quantum counterparts of classical bits, namely quantum bits (or qubits). Quantum computers make use of the quantum mechanical properties of nature, as such qubits have different properties than their classical counterparts and the operations we can apply to them are also different. Where a bit can be either $0$ or $1$, a qubit can be in the state $\ket{0}$ or $\ket{1}$. 
A single qubit can hold any value of the form $\ket{\psi}$ (pronounced `ket psi')
\begin{equation}
    \ket{\psi}=\alpha_0\ket{0}+\alpha_1\ket{1},\quad \alpha_0,\alpha_1\in\mathbb{C},\quad |\alpha_0|^2+|\alpha_1|^2=1.
    \label{eq:qubit}
\end{equation}
the $\alpha_i$ are referred to as the probability amplitudes. When measuring a general quantum state $\ket{\phi}$ we will find the state $\ket{0}$ or $\ket{1}$ with probabilities equal to $|\alpha_0|^2$ and $|\alpha_1|^2$ respectively. 

Equivalently we can interpret the computational basis states $\ket{0}$ and $\ket{1}$ as vectors in 2-dimensional Hilbert space, this gives
\begin{equation}
    \ket{\psi}
    =
    \alpha_0\ket{0}+\alpha_1\ket{1}
    =
    \alpha_0
    \begin{bmatrix}
    1\\0
    \end{bmatrix}
    +\alpha_1
    \begin{bmatrix}
    0\\1
    \end{bmatrix}
    =
    \begin{bmatrix}
    \alpha_0\\\alpha_1
    \end{bmatrix}.
    \label{eq:qubitBasisCoordinates}
\end{equation}

We can extend this to $n$ qubits by writing the most general quantum state as
\begin{equation}
    \ket{\boldsymbol{\psi}}
    =
    \sum_{b\in\{0,1\}^n}
    \alpha_{b}\ket{b}
    =
    \sum_{k=0}^{2^n-1}
    \alpha_{k}\ket{k},
    \quad
    \alpha_k\in\mathbb{C},
    \quad
    \sum_{k=0}^{2^n-1}
    |\alpha_{k}|^2=1.
    \label{eq:qubit_register}
\end{equation}
where in the first equation we switched from the binary labeling of the basis states to their associated natural number. 
As before an $n$-qubit state $\ket{\psi}$ can be interpreted as a normalised vector in $2^n$ dimensional complex space. The vector has size $2^n$ as there are $2^n$ possible basis states, one for each possible combination of separate bit states.

\subsection{Quantum gates}
Quantum gates are the elementary operations with which the qubit states can be manipulated. All quantum gates are represented mathematically as unitary matrices, that is, a complex-valued square matrix $U\in\mathbb{C}^{2^n\times 2^n}$ such that $UU^\dagger=I$, where superscript $\dagger$ denotes the conjugate transpose\footnote{The superscript $\dagger$ to respresent the conjugate transpose is most commonly used in quantum computation and is traditionally a physics notation, mathematicians usually prefer the superscript $*$.  }. The unitary nature of all quantum gates implies that all quantum gates must be reversible. It also leads to the quantum no-cloning theorem, which shows that arbitrary quantum states cannot be copied \cite{mikeandike}.

\subsection{Measurement}
\label{sec:measurement}

Let us come back to the measurement process which is the primary way to retrieve information from a quantum computation. As it is not possible to read out the probability amplitudes $\alpha_k\in\mathbb{C}$ of a quantum state, we have to deduce them indirectly by letting the quantum state collapse to a classical binary state and repeating this process many times to be able to reconstruct a probability distribution over all possible outcomes.

Let $\ket{\psi}$ be as defined in Equation \eqref{eq:qubit_register} and let $\ket{i}$ represent the $i$-th basis state of the computational basis. The probability of measuring $\ket{\psi}$ in the state $\ket{i}$ is given by
\begin{equation}
    P_{\ket{\psi}}(\ket{i}) = |\braket{i|\psi}|^2 = |\alpha_i|^2.
\end{equation}

One of the special properties of quantum computation is that once we measure a quantum state $\ket{\psi}$ in the computational basis and find the state $\ket{i}$ upon measurement, the quantum state has collapsed to $\ket{i}$. This means that after our measurement the quantum state will collapse to be
\begin{equation}
    \ket{\psi} = \frac{\alpha_i}{\|\alpha_i\|_2}\ket{i}.
\end{equation}

\subsection{Entanglement}
We will explain the concept of entanglement using an example. Let us consider one particular quantum state, the Bell state, which is defined on two qubits
\begin{equation}
    \ket{\boldsymbol{\psi}}
    =
    \frac{1}{\sqrt{2}}\ket{00}+\frac{1}{\sqrt{2}}\ket{11}.
    \label{eq:firstBellState}
\end{equation}
If we stick to the interpretation of measurement as the act of forcing the quantum state into one of the computational basis states the first Bell state collapses to $\ket{00}$ and $\ket{11}$ with probability $\left(1/\sqrt{2}\right)^2=1/2$ each. The other two possible states, $\ket{01}$ and $\ket{10}$, never occur. Knowing this, it suffices to measure only one of the two qubits to immediately know what will be the measurement outcome of the other. This special property in which qubits are no longer independent from each other is termed entanglement of qubits. 

\subsection{Superposition}
Consider a quantum state 
\begin{equation}
    \ket{\psi} = \frac{1}{N} \sum_{k=0}^{N-1} \ket{k}.
\end{equation}
This state $\ket{\psi}$ is in a so-called superposition of basis states. To be more precise, it is in an equal superposition of the $N$ basis states. 
Now, if we apply a quantum operation $U$ to this state we get
\begin{equation}
    U\ket{\psi} = \frac{1}{N} \sum_{k=0}^{N-1} U \ket{k}.
\end{equation} 
Which means we can apply the operation $U$ to all basis states in a single operation. The ability to apply an operation to multiple basis states at once is sometimes referred to as quantum parallelism and it is one of the main features of quantum computing that we wish to exploit.
Notice that due to the measurement properties of quantum computing we are, unfortunately, not able to directly retrieve the values $U\ket{k}$.

\section{Glossary}\label{app:glossary}

\begin{center}
\begin{tabular}{ | c || c | c |}
\hline
 $a_{l,i}$ & \makecell{The ancilla qubit representing \\ whether a particle is higher than \\ the lowest point of a wall $a$} & \\
\hline
 $a_{o,i}$ & \makecell{The ancilla qubit associated to the\\reflection step for\\the $i$-th dimension.} & \\
\hline
 $a_{u,i}$ & \makecell{The ancilla qubit representing \\ whether a particle is lower than \\ the highest point of a wall $a$} & \\
 \hline
$a_{v,i}$ & \makecell{The ancilla qubit representing\\whether or not a particle\\should take a step in the\\$i$-th dimension in the current timestep.} &\\
    \hline
      \makecell{C$^p$NOT gate } & \makecell{Multi-controlled NOT gate \\ controlled by $p$ qubits. } & \phantom{$n_v = \sum_{i=1}^d n_{v_i}$} \\ 
 \hline
    $c_{\text{rel}}$ & \makecell{The relative speed of \\a velocity vector $\mathbf{u}$\\in the separate dimensions. } & $c_\text{rel} = \max_{i,j\in d} \frac{|u^{(i)}|}{|u^{(j)}|}$\\
 \hline
  $g_j$ & \makecell{Label for the $j$-th qubit representing\\the position of the particle.}  & \\ \hline
  $g^i_l$ & \makecell{The $l$-th qubit representing\\the position of the particle\\in the $i$-th dimension.}  & \\ 
  \hline
   $n_a$ & \makecell{The total number of \\ ancilla qubits required \\ for the implementation.} & $n_a = 4d -2$\\
 \hline
  $n_g$ & \makecell{The total number of\\qubits required to\\express the grid.} &  $n_g = \sum_i^{d} n_{g_i}$ \\
 \hline
 $n_{g_i}$ & \makecell{The number of qubits\\required to express the\\location of the $i$-th dimension.} & \\
 \hline
  $n_{g_\text{max}}$ & \makecell{The maximum number of qubits\\required to express\\the location in any dimension.} & $n_{g_\text{max}}= \max_{i\in \{1,\dots,d\}} n_{g_i} $ \\
 \hline
 $n_v$ & \makecell{The total number of qubits\\required to express the\\velocity vector.} & $n_v = \sum_{i=1}^d n_{v_i}$ \\
  \hline
 $n_{v_i}$ & \makecell{The number of qubits\\required to express\\the velocity of the $i$-th dimension.} & \\
 \hline
   $n_{v_\text{max}}$ & \makecell{The maximum number of qubits\\required to express\\the velocity in any dimension.} & $n_{v_\text{max}}= \max_{i\in \{1,\dots,d\}} n_{v_i} $ \\
    \hline

 \end{tabular}
 
 \newpage
 
 \begin{tabular}{ | c || c | c |}
 \hline
   $N_{v_i}$ & \makecell{The number of distinct \\ velocities in dimension $i$} & $N_{v_i} = |\mathcal{U}^{(i)}|$ \\
 \hline
  $N_t$ & \makecell{Total number of timesteps} & \\
 \hline
   $N_{v_i}$ & \makecell{The number of distinct \\ velocities in all dimensions combined} & $N_{v} = |\mathcal{U}|$ \\
 \hline
  $n^v_t$ & \makecell{The number of velocities \\ with distinct absolute values $|u_i|$ \\ for which we take a step \\ in timestep $t$ } & $n_t^v \leq \frac{N_v}{2}$ \\
 \hline
 \makecell{$n_w$ } & \makecell{The number of walls \\ present in the problem.\\ } & \phantom{$n_v = \sum_{i=1}^d n_{v_i}$} \\ 
 \hline
  $T$ & Total time for which the simulation is run. & $\sum_{i=0}^{N_t-1}\Delta t_i$  \\ 
  \hline
    $\mathbf{u}$ & \makecell{Velocity vector.}  &  $\mathbf{u}=\left ( u^{(0)}, u^{(1)}, \dots, u^{(d)} \right )$  \\ 
 \hline
   $\mathcal{U}$ & \makecell{Set of possible speeds\\ in all dimensions combined.} & $\mathcal{U} = \bigcup_{i=1}^d  \mathcal{U}^{(i)}$\\
 \hline
   $u^{(i)}$ & \makecell{Speed of a particle \\ in dimension $i$.}  &   \\ 
 \hline
  $\mathcal{U}^{(i)}$ & \makecell{Set of possible speeds\\ in dimension $i$.} & $\mathcal{U}^{(i)} = \{u_1, \dots , u_{N_v} \}$\\
 \hline
  $\mathcal{U}_o$ & \makecell{Ordered list of possible \\ speeds .} & $\mathcal{U} = [-u_{\max}, \dots , u_{\max} ]$\\
 \hline
    $U_{P,+}$ & \makecell{Quantum operation consisting \\of one phase shift gate \\per qubit, designed to \\increase the position of \\the basis state by 1 \\in Fourier space.}  &    \\ 
 \hline
   $U_{P,-}$& \makecell{Quantum operation consisting \\of one phase shift gate \\per qubit, designed to \\decrease the position of \\the basis state by 1 \\in Fourier space.}  &   \\ 
 \hline
    $v_{\text{dir},i}$ & \makecell{The qubit expressing the\\direction of the velocity of\\the particle in the $i$-th direction.} &\\
 \hline
 $v_j$ & \makecell{Label for the $j$-th qubit representing\\the velocity of the particle.}  & \\
    \hline
 $v^i_{l}$ & \makecell{The $l$-th qubit representing\\the speed of the particle\\in the $i$-th dimension.}  & \\
  \hline
    $\mathbf{x}$ & \makecell{Vector encoding a position in space.}  &  $\mathbf{x}= \left ( x^{(0)}, x^{(1)}, \dots, u^{(d)} \right )$  \\ 
 \hline
   $x^{(i)}$ & \makecell{Position of a particle \\ in dimension $i$.}  &   \\ 
 \hline

\end{tabular}

\newpage
 
\begin{tabular}{ | c || c | c |}
    \hline
 x-wall & \makecell{Wall of which the normal points in the \\ x-direction.} & \phantom{$\mathbf{x}= \left ( x^{(0)}, x^{(1)}, \dots, u^{(d)} \right )$ }\\
 \hline
 y-wall & \makecell{Wall of which the normal points in the \\ y-direction.} & \\
 \hline
 z-wall & \makecell{Wall of which the normal points in the \\ z-direction.} &\\
 \hline

     $\Delta t_m$  & \makecell{Size of timestep $m$.} &  \\  
 \hline
\end{tabular}

\end{center}

\end{document}